\documentclass[twocolumn]{aastex631}

\newcommand\textline[3][t]{%
  \par\smallskip\noindent\parbox[#1]{.156\textwidth}{\raggedright#2}%
  \parbox[#1]{.156\textwidth}{\raggedleft\texttt{#3}}\par\smallskip%
}

\shorttitle{GALFIT-ing AGN Host Galaxies in COSMOS: \textit{HST} vs. Subaru}
\shortauthors{Dewsnap et al.}

\graphicspath{{./}{figures/}}

\begin{document}

\title{GALFIT-ing AGN Host Galaxies in COSMOS: \textit{HST} vs. Subaru}

\correspondingauthor{Corresponding Author}
\email{cdewsnap@uwo.com}

\author[0000-0003-1225-3462]{Callum Dewsnap}
\affiliation{Department of Physics \& Astronomy, The University of Western Ontario, London, ON, Canada}

\author[0000-0003-2767-0090]{Pauline Barmby}
\affiliation{Department of Physics \& Astronomy, The University of Western Ontario, London, ON, Canada}
\affiliation{Institute for Earth \& Space Exploration, The University of Western Ontario, London, ON, Canada}

\author[0000-0001-6217-8101]{Sarah C. Gallagher}
\affiliation{Department of Physics \& Astronomy, The University of Western Ontario, London, ON, Canada}
\affiliation{Institute for Earth \& Space Exploration, The University of Western Ontario, London, ON, Canada}

\author[0000-0002-0745-9792]{C. Megan Urry}
\affiliation{Yale Center for Astronomy and Astrophysics, Department of Physics, Yale University, New Haven, CT, USA}

\author[0000-0002-2525-9647]{Aritra Ghosh}
\affiliation{Yale Center for Astronomy and Astrophysics, Department of Astronomy, Yale University, New Haven, CT, USA}

\author[0000-0003-2284-8603]{Meredith C. Powell}
\affiliation{Kavli Institute of Particle Astrophysics and Cosmology, Stanford University, Stanford, CA, USA}

\begin{abstract}

The COSMOS field has been extensively observed by most major telescopes, including \textit{Chandra}, \textit{HST}, and Subaru. \textit{HST} imaging boasts very high spatial resolution and is used extensively in morphological studies of distant galaxies. Subaru provides lower spatial resolution imaging than \textit{HST} but a substantially wider field of view with greater sensitivity. Both telescopes provide near-infrared imaging of COSMOS. Successful morphological fitting of Subaru data would allow us to measure morphologies of over $10^4$ known active galactic nucleus (AGN) hosts, accessible through Subaru wide-field surveys, currently not covered by \textit{HST}. For 4016 AGN between $0.03<z<6.5$, we study the morphology of their galaxy hosts using GALFIT, fitting components representing the AGN and host galaxy simultaneously using the \textit{i}-band imaging from both \textit{HST} and Subaru. Comparing the fits for the differing telescope spatial resolutions and image signal-to-noise ratios, we identify parameter regimes for which there is strong disagreement between distributions of fitted parameters for \textit{HST} and Subaru. In particular, the S\'ersic index values strongly disagree between the two sets of data, including sources at lower redshifts. In contrast, the measured magnitude and radius parameters show reasonable agreement. Additionally, large variations in the S\'ersic index have little effect on the $\chi^2_\nu$ of each fit whereas variations in other parameters have a more significant effect. These results indicate that the S\'ersic index distributions of high-redshift galaxies that host AGN imaged at ground-based spatial resolution are not reliable indicators of galaxy type, and should be interpreted with caution.

\end{abstract}


\section{Introduction} \label{sec:intro}

The properties of the galaxies that host active galactic nuclei (AGN) offer clues to the conditions that enable accretion onto supermassive black holes (SMBHs). Because galaxy shape is linked to various mechanisms of galaxy evolution, measuring morphologies is an important step in characterizing AGN host galaxies. Additionally, by using AGN samples over a large redshift range, we are offered a unique view into the evolution of SMBHs over cosmic time frames, as the actively accreting matter is feeding the growth of the black holes.

X-ray surveys are used extensively to detect AGN, in part due to the lack of contamination from other strong sources of continuum emission such as star formation \citep{Civano2012}. IR surveys are also fairly common, but can run into issues with contamination by, for example, star-forming galaxies particularly in specific redshift ranges \citep{Donley2012}. Although each method has its own benefits and drawbacks, different classes of AGN are more readily detected by different methods, and thus to create a complete collection of data we must utilize multi-wavelength surveys. This allows for a more complete understanding of the AGN contained within the survey \citep{Hickox2009}. Using these multi-wavelength surveys, an AGN could be detected in the X-ray, then have the corresponding counterparts in the IR and optical detected for follow-up study, as was done in \citet{Civano2012}.

An example of an extragalactic multi-wavelength survey is the Cosmic Evolution Survey (COSMOS), a ${\sim} 2 \ \mathrm{deg}^2$ region which has been observed by essentially every major space telescope as well as many ground-based telescopes. The telescopes include the \textit{Hubble Space Telescope} (\textit{HST}), Subaru Telescope, Canada-France-Hawaii Telescope (CFHT), \textit{Chandra X-ray Observatory}, \textit{XMM-Newton}, Keck, \textit{Spitzer}, etc. \citet{Scoville2007} provide a summary of COSMOS observations and includes estimates for the number of extragalactic objects within. These observations provide highly beneficial layers of depth for a very wide range of wavelengths, a necessary factor in understanding AGN of all types and their hosts.

In this project, we characterize the morphological nature of galaxies at redshift $0.03 < z < 6.5$ that host AGN and compare the results for higher- and lower-angular resolution data. We do this by using two sets of imaging data containing a large sample of active galaxies in common and measure a set of morphological parameters of each independently. We then compare and discuss the trends which occur between the two sets of data. Agreement between the two sets of data would unlock a large sample of data which we could not fit previously, as the lower angular resolution, ground-based data cover a much larger area. These larger sample sizes would provide a much more comprehensive understanding regarding the link between AGN and their host galaxies.

\section{Images and Data} \label{sec:data}

Spatial resolution is of the utmost importance when studying morphology. Historically, the vast majority of morphological-fitting studies of distant galaxies use \textit{HST} imaging. This is especially true when studying AGN, as \textit{HST} provides the necessary high angular resolution and stable, well-characterized point-spread function (PSF) to disentangle the point-like AGN and its host galaxy. \textit{HST} thus provides a clearer distinction between the central bulge and point-source, an important factor to consider when trying to account for the flux of each component individually. These benefits do come at a cost, however, as in morphological studies such as these we require deep, high resolution imaging for a relatively large region of the sky to obtain sufficiently large samples. The \textit{HST} Advanced Camera for Surveys' Wide Field Channel (ACS/WFC), however, has a relatively small field of view (FOV), at only $202 \times 202 \ \mathrm{arcsec}^2$ which is nonideal for use in large surveys.

The COSMOS field is unique in the fact that there exists full contiguous \textit{HST} ACS/WFC coverage of the region with a median exposure depth of 2028 seconds (a full \textit{HST} orbit). Thus, we can use these data as a testing ground for comparisons between lower-resolution observations taken by telescopes more apt for large-area surveys. The Subaru Telescope's Hyper Suprime-Cam (HSC), a camera for the ground-based $8$-metre telescope, has a poorer spatial resolution than that of \textit{HST} (with seeing-dominated spatial resolution ${\sim}6.3\times$ larger), while covering a much larger FOV of $90 \times 90 \ \mathrm{arcmin}^2$ \citep{Miyazaki2012}. In general, large, ground-based telescopes will be less sensitive to faint point-sources than \textit{HST}, but can be more sensitive to larger, extended sources such as galaxy disks. The COSMOS field has been completely imaged by Subaru HSC, allowing for an easy comparison of the difference in morphological parameters measured. Subaru HSC also provides excellent coverage of other large regions that have AGN catalogs but no \textit{HST} coverage. This makes the comparison very useful as, if our method is validated, we gain access to a large sample of Subaru HSC-quality data previously considered to be of uncertain reliability.

\begin{figure}[t]
\centering
\includegraphics[width=0.47\textwidth]{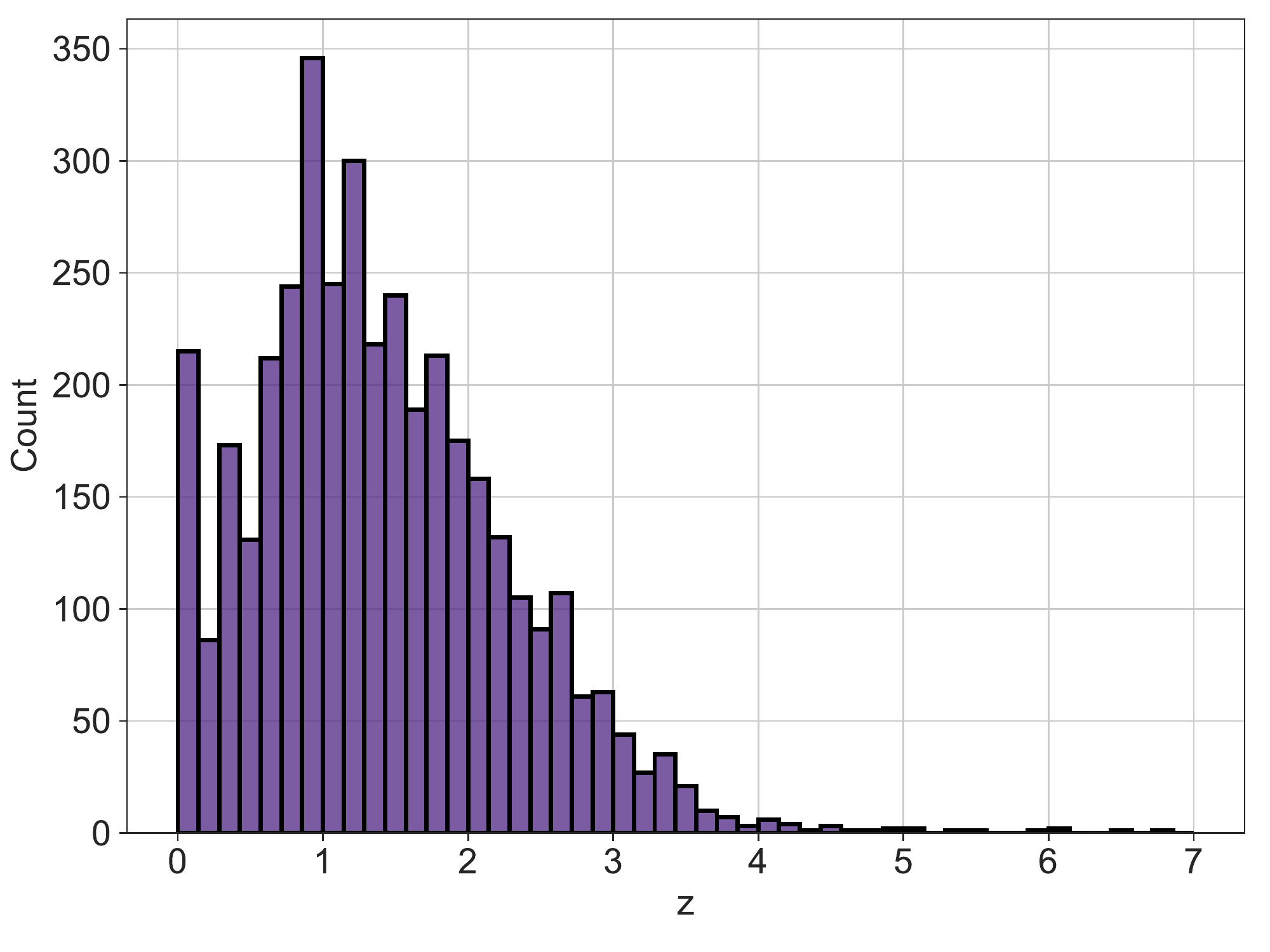}
\caption{Redshift distribution of sources from the \citet{Marchesi2016} X-ray AGN catalog.}
\label{fig:redshift_dist}
\end{figure}

The abundance of multiwave observations in the COSMOS field makes it an invaluable tool in understanding AGN. Due to the high sensitivity and resolution of the data, COSMOS is an excellent source of high redshift objects, with sources as high as $z \sim 6.5$ \citep{Scoville2007}. In this study, we use the AGN catalog of \citet{Marchesi2016} who identified optical and IR counterparts of the \textit{Chandra} COSMOS-Legacy Survey \citep{Elvis2009, Civano2015}. This catalog provides us with 4016 X-ray sources, 97\% of which have an optical/IR counterpart as well as a photometric redshift measurement. About ${\sim} 54\%$ of these sources have a spectroscopic redshift measurement. This catalog also provides numerous measured X-ray and optical properties for each source. Figure \ref{fig:redshift_dist} shows the redshift distribution of the AGN within the catalog.

In this study, we use high-resolution \textit{HST} ACS/WFC imaging of the entire COSMOS field \citep{Koekemoer2007, Massey2010} with a pixel scale of $0.030"/\mathrm{pixel}$. The limiting point-source depth of the \textit{HST} imaging is a magnitude of $F814(AB)=27.2$ ($5\sigma$). For each of the 4016 sources, a cutout is extracted. These cutouts range in size from 8--30 arcsec (267--1000 pixels). These cutouts are fit for the \textit{HST} portion of this study. The PSF used for the \textit{HST} fits was created using TinyTim \citep{Krist2011} and has a FWHM of 0.095{\arcsec}.

We used the AGN catalog positions to identify the cutout locations for the COSMOS field using the Subaru HSC imaging from the Hyper Suprime-Cam Subaru Strategic Program (HSC-SSP) \citep{Aihara2018, Aihara2019}. This program provides a Wide layer of data covering an area of ${\sim} 300 \ \mathrm{deg}^2$. Within this area, there exists ${\sim} 27 \ \mathrm{deg}^2$ and ${\sim} 3.5 \ \mathrm{deg}^2$ of complete Deep and UltraDeep data, respectively. These Deep/UltraDeep data are jointly processed and cover the entire COSMOS field with a $5\sigma$ point-source depth magnitude of 27.1 in the \textit{i}-band. In terms of point-source depth, this is shallower than the \textit{HST} observations, however Subaru HSC provides deeper imaging for extended sources at only ${\sim} 0.3$ magnitudes fainter within a 2{\arcsec} diameter aperture than its point-source depth. The Subaru HSC imaging has a pixel scale of $0.168"/\mathrm{pixel}$. Cutouts are taken from these data using the cutout tool provided by \citet{Aihara2019}. \citet{Aihara2019} provide a PSF picker with which we selected the corresponding PSF for each cutout. The median PSF FWHM of the Subaru HSC \textit{i}-band data is 0.66{\arcsec}.

It is important to contextualize the cost of collecting the two sets of data. While \textit{HST} has superior spatial resolution, it required 583 orbits of \textit{HST} observation, each with a 2028 second exposure, to fully cover $1.64 \ \mathrm{deg}^2$ of the COSMOS field. This means that observations took approximately two weeks of observing time, forcing the observations to take place over the course of two years. In the Subaru HSC imaging done by \citet{Aihara2019}, the Deep set of imaging covered a total of $27 \ \mathrm{deg}^2$ over 10 exposures, taking 2.1 hours. The UltraDeep imaging covered an area of $3.5 \ \mathrm{deg}^2$ over 20 exposures, taking 14 hours. Given how substantially less time is required for these observations, it took only approximately ten days as opposed to the two years of \textit{HST}. Even with ground-based telescopes such as Subaru being held within the limitations of weather and Earth's rotation, the advantage for large-scale surveys is clear, and the ability to perform morphological fits despite the lower spatial resolution would be extremely valuable. In addition, the Subaru HSC Wide set of imaging covers ${\sim}1400 \ \mathrm{deg}^2$ containing millions of galaxies, including many AGN hosts. This sample is invaluable when comparing to the 4016 sources in this study.

\section{Fitting Process} \label{sec:process}

In order to characterize the morphology of a galaxy, we can model the surface brightness profile using a mathematical function. By fitting this function to the surface brightness profile, we can measure certain properties about the galaxy, such as the effective radius or magnitude. 2D fitting fits brightness profiles directly to an image of a galaxy. This fitting process involves the convolution of the model with a PSF, thus accounting for image smearing. Another primary benefit of 2D fitting is the ease of visually checking the results of the fit. Because the model is fit directly to the image of the galaxy, one can simply create a residual image by taking the difference between the observed and model images for a simple check of fit quality. There are a number of different 2D fitting software packages, for example GALFIT \citep{Peng2002, Peng2010}, GIM2D \citep{Simard1998}, and BUDDA \citep{deSouza2004}. The versatility of these software packages allows them to be used for many different studies of galaxy morphology for a wide range of data \citep{Haussler2007, Gabor2009, Sheth2010, vanderWel2012, Bottrell2019, Li2021}.

The most commonly used brightness profile is the S\'ersic profile \citep{Sersic1963, Sersic1968}. This profile is defined as
\begin{equation}
    \Sigma (r) = \Sigma_e \exp{\left[ -\kappa_n \left( \left( \frac{r}{r_e} \right)^{\frac{1}{n}} -1 \right) \right]},
\end{equation}
where $\Sigma_e$ is the pixel surface brightness at the half-light radius $r_e$ (defined to be the radius of the isophote containing half of the luminosity of the galaxy), $n$ a parameter called the S\'ersic index, and $\kappa_n$ a variable dependent on $n$ defined by
\begin{equation}
    \gamma (2n;\kappa_n) = \frac{1}{2} \Gamma (2n),
\end{equation}
where $\Gamma$ and $\gamma$ are the Gamma and lower incomplete Gamma functions, respectively. The S\'ersic profile with $n=1$ is identical to the exponential disk profile used to model spiral galaxies \citep{Freeman1970}. Similarly, the value $n=4$ gives the de Vaucouleurs profile historically used to model many elliptical galaxies \citep{deVaucouleurs1948, deVaucouleurs1976, deVaucouleurs1991}. The total flux of the source can be calculated by integrating $\Sigma(r)$ out to $r=\infty$, resulting in the expression
\begin{equation} \label{eq:flux}
    F_\mathrm{total} = 2 \pi r_e^2 \Sigma_e \mathrm{e}^{\kappa_n} n \kappa_n^{-2n} \Gamma(2n) q / R(C_0;m),
\end{equation}
where $q$ is the axis ratio and $R(C_0;m)$ is a geometric correction factor. This correction factor is typically 1 and is related to optional Fourier modes and a diskiness/boxiness factor which are used only for complicated fits of nearby galaxies \citep{Peng2010}.

Depending on the complexity of the fit, there are multiple common techniques to apply the above profiles. If a galaxy is imaged at sufficiently high resolution, the galaxy can be modelled by simultaneously fitting components for the disk and bulge, possibly using either an exponential disk alongside a de Vaucouleurs profile, or simply two S\'ersic profiles. It would also be possible to add additional profiles to model other phenomena, for example a bar or a ring. If the galaxy has small angular size compared to the telescope resolution, a single S\'ersic index is often able to fit the source with sufficient accurately.

When performing morphological fits on large-scale survey data, it is important to account for the central point-source seen in active galaxies, even in studies which are not focused on AGN. A single galaxy brightness profile cannot accurately account for the galaxy and the AGN simultaneously, and thus the AGN must be treated separately. Because an AGN appears as a distinct point-source, the most common method to fit AGN is to simultaneously fit a central point-source alongside the typical galaxy components. This method is standard among morphology studies using either 1D or 2D fitting, and has been thoroughly tested for its viability \citep{Kim2008, Simmons2008}.

In order to perform the morphological fits, we apply the 2D fitting software GALFIT \citep{Peng2002, Peng2010}. We selected GALFIT due to its prevalence in literature as well as its ability to fit any number of components simultaneously. GALFIT uses a nonlinear least-squares algorithm which applies the Levenberg-Marquardt technique, an algorithm which is among the most efficient for large parameter spaces. In each step of a fit, GALFIT calculates a $\chi^2$ value and computes how it should alter the morphological parameters in order to minimize the $\chi^2$. It continues to iterate until either the $\chi^2$ value converges or a maximum number of iterations is reached. A basic value with which we can measure the quality of the fit is the reduced $\chi^2$, called $\chi_\nu^2$, defined as
\begin{equation} \label{eq:chi2nu}
    \chi^2_\nu = \frac{1}{N_\mathrm{dof}} \sum_{x=1}^{n_x} \sum_{y=1}^{n_y} \frac{\left( f_\mathrm{data}(x,y) - f_\mathrm{model}(x,y) \right)^2}{\sigma(x,y)^2},
\end{equation}
with
\begin{equation}
    f_\mathrm{model}(x,y) = \sum_{\nu=1}^{m} f_\nu(x,y;\alpha_1,\ldots,\alpha_n).
\end{equation}
Here we have $n_x$ and $n_y$ the dimensions of the image, $f_\mathrm{data}(x,y)$ the flux measured at the point $(x,y)$, and $f_\mathrm{model}(x,y)$ the sum of each component function $f_\nu(x,y;\alpha_1,\ldots,\alpha_n)$, with $(\alpha_1,\ldots,\alpha_n)$ the free parameters of the fit. For each component function, $\nu$ represents the component number with $m$ the total number of components. $N_\mathrm{dof}$ is the number of degrees of freedom. This term is defined as the difference between the number of pixels in the image and the number of free parameters in the model. The function $\sigma(x,y)$ is the Poisson error at each point of the image. This is computed through the use of a sigma image which can either be input by the user or calculated by GALFIT using the GAIN, EXPTIME, and NCOMBINE headers from the cutout. \citet{Gabor2009} find that the choice of sigma image leads to small uncertainties relative to those introduced by the PSF and other effects.

GALFIT allows for the use of many different brightness profiles, including the S\'ersic profile. GALFIT is able to convolve each profile with a user-input PSF in order to accurately model the image spread seen in the observations. Alongside this, GALFIT is able to fit a point-source (PS) component alongside any number of other components. This PS component is also convolved with the PSF and thus simply appears in the image as the PSF.

\begin{figure*}[t]
\centering
\includegraphics[width=\textwidth]{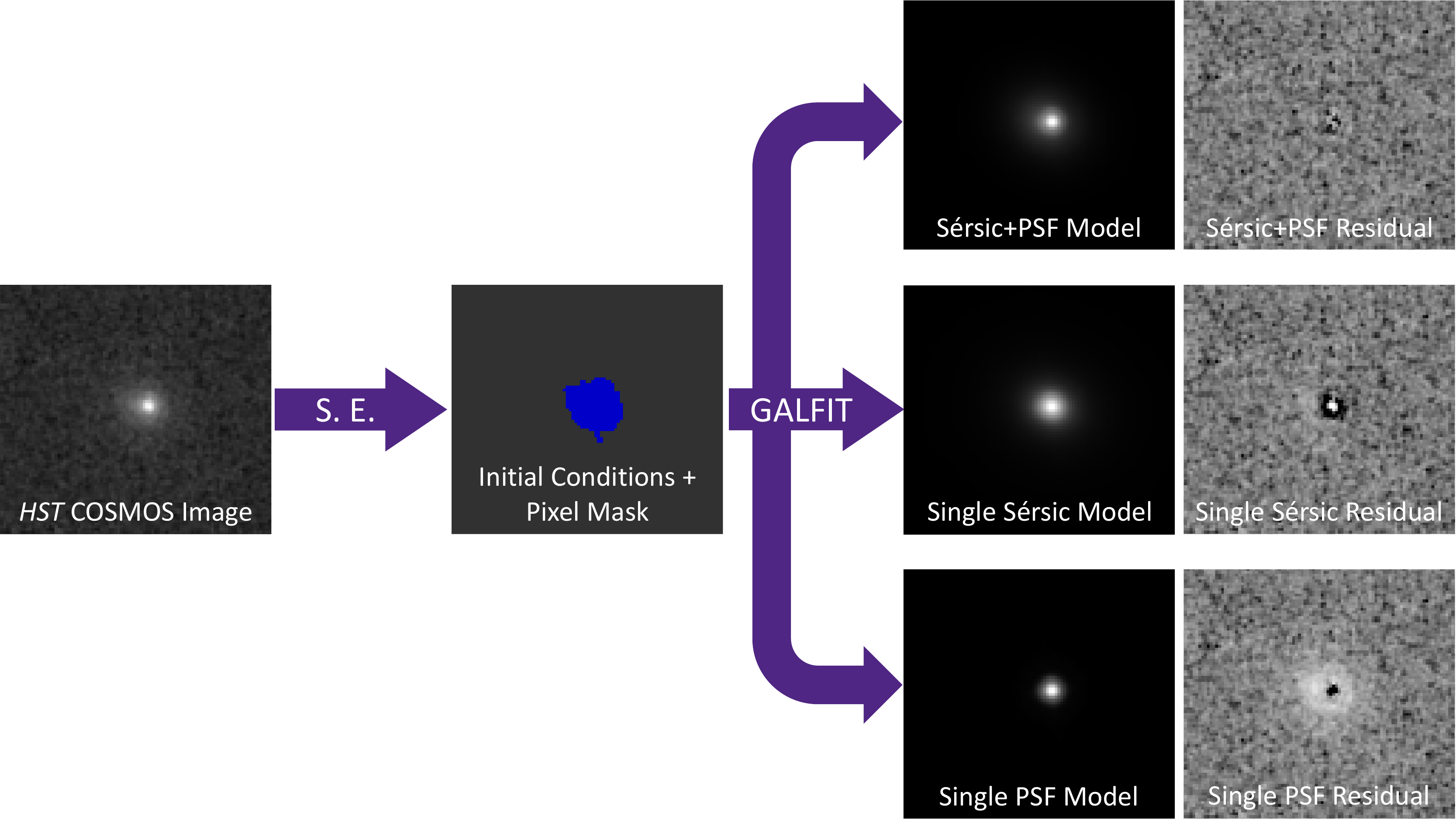}
\caption{Example of the fitting process showing typical results for each of the three fits. The source is identified as `cid\_380' by the source catalog \citep{Marchesi2016}. Each cutout has an angular width of 3.03{\arcsec}. S.E. represents the Source Extractor step in which we determine the initial conditions as well as the pixel mask. In this case, there are no neighboring sources to be masked out. The single S\'ersic fit fails near the the central point-source and the single PS fit fails for the extended host galaxy. The S\'ersic+PS fit best accounts for both the host galaxy and AGN contributions.}
\label{fig:goldilocks}
\end{figure*}

There are a number of possible ways in which we can fit the surface brightness profiles of the set of galaxies. In this study, we apply three different fits to each galaxy in the catalog: one fit with a single S\'ersic profile component, one with only a single point-source component, and one with both a S\'ersic and a PS component. In most cases, the AGN will have a significant contribution to its host galaxy's brightness profile. This means that, in general, the S\'ersic+PS fit will best account for the source, as the single S\'ersic fit will fail near the central point-source and the single PS fit will fail for the extended galaxy. If the central point-source is faint relative to the galaxy, it can be more difficult to differentiate between the galaxy bulge and the AGN. If the bulge is small relative to the image resolution (i.e., the bulge is comparable in size to the image PSF), then the S\'ersic+PS fit may over-fit the region near the point-source leaving the single S\'ersic fit as the best fit. Even if the bulge is not comparable in size to the PSF, a faint point-source will result in a more uncertain determination of the host galaxy and AGN properties. An example of each of the three fits is given in Figure \ref{fig:goldilocks}. Due to the redshift distribution of the galaxies in our sample, it is possible that the galaxy and AGN appear as a single point-source. This is due to both the smaller angular size of the observed galaxies and selection bias; only more luminous AGN are detected at higher redshift and thus are more likely to dominate their host galaxy. In this case, we again may not be able to differentiate the galaxy from the AGN, resulting in the single PS fit (or possibly the single S\'ersic fit with high $n$) providing the best fit. The S\'ersic+PS fit is most valuable, as it is the only fit which is able to potentially provide morphological parameters of the host galaxy as well as separate flux measurements of the galaxy and AGN.

The S\'ersic profile has a number of free parameters to be fit and provides useful derived galaxy properties. Outputs include the position on the image $\left(x_0,y_0\right)$, the total integrated magnitude $m_\mathrm{host}$, the effective (half-light) radius $r_e$, the S\'ersic index $n$, the axis ratio $q$ (defined as $b/a$, with $a$ and $b$ the semi-major and semi-minor axes), and the position angle $\theta_\mathrm{P.A.}$ (defined with the positive $y$ direction as $0^{\circ}$ increasing counterclockwise). The PS component provides a position $\left(x_\mathrm{PS},y_\mathrm{PS}\right)$ and a total integrated magnitude $m_\mathrm{PS}$. In order to define an AB magnitude, we must calculate the corresponding zero-point. For an \textit{HST} ACS image, this is a relatively simple process as outlined by the work of \citet{Bohlin2016}. The Subaru data provide the zero-point in units of flux in the image header for straightforward conversion into a magnitude.

GALFIT allows for constraints to be applied to the free parameters of a fit. The constraints are selected such that the results remain physical in cases where the solution does not converge and in order to prevent GALFIT from crashing if extreme values are reached. Note that the constraints are not intended to limit the range of variation in the fitting process. In order to apply the $\chi^2$ minimization, the parameters must be allowed to vary outside of physically expected results. The constraints are summarized in Table \ref{tab:constraints}. The constraints we selected are fairly standard throughout similar studies \citep{Simmons2008, Gabor2009, vanderWel2012, Powell2017, Bottrell2019, Ishino2020, Li2021}. We experimented with different sets of constraints and found that the vast majority of fits see little change in the best fit parameters as the constraints are tightened or loosened, a result in agreement with \citet{Gabor2009}. If the constraints are loosened, fits which reach the boundary in the more strict case tend to also reach to the boundary in the less strict case.

\begin{deluxetable*}{lccc}
\tablecaption{The constraints applied to each fit. \label{tab:constraints}}
\tablehead{
\colhead{Parameter Name} & \colhead{S\'ersic+PS Fit} & \colhead{S\'ersic Fit} & \colhead{PS Fit}}
\startdata
    S\'ersic Index $\left(n\right)$ & $0.5<n<8$ & $0.5<n<8$ & N/A \\
    Host Magnitude $^\mathrm{a}$ & $\left|m_\mathrm{host} - m_\mathrm{init}\right| \leq 2.5$ & $\left|m_\mathrm{host} - m_\mathrm{init}\right| \leq 2.5$ & N/A \\
    Half-Light Radius [arcsec] & $0<r_e<9$ & $0<r_e<9$ & N/A \\
    $x_\mathrm{host}$ Position [arcsec] $^\mathrm{b}$ & $\left|x_\mathrm{host} - x_\mathrm{init}\right| \leq 0.6$ & $\left|x_\mathrm{host} - x_\mathrm{init}\right| \leq 0.6$ & N/A \\
    $y_\mathrm{host}$ Position [arcsec] $^\mathrm{b}$ & $\left|y_\mathrm{host} - y_\mathrm{init}\right| \leq 0.6$ & $\left|y_\mathrm{host} - y_\mathrm{init}\right| \leq 0.6$ & N/A \\
    Point-Source Magnitude $^\mathrm{c}$ & $\left|m_\mathrm{PS} - m_\mathrm{init}\right| \leq 7.5$ & N/A & $\left|m_\mathrm{PS} - m_\mathrm{init}\right| \leq 7.5$ \\
    $x_\mathrm{PS}$ Position [arcsec] $^\mathrm{d}$ & $\left|x_\mathrm{PS} - x_\mathrm{host}\right| \leq 0.15$ & N/A & $\left|x_\mathrm{host} - x_\mathrm{init}\right| \leq 0.6$ \\
    $y_\mathrm{PS}$ Position [arcsec] $^\mathrm{d}$ & $\left|y_\mathrm{PS} - y_\mathrm{host}\right| \leq 0.15$ & N/A & $\left|y_\mathrm{host} - y_\mathrm{init}\right| \leq 0.6$ \\
\enddata
\tablecomments{
    $^\mathrm{a}$ The host magnitude is constrained to be within 2.5 of the initial host magnitude input.\\
    $^\mathrm{b}$ The position of the host galaxy is constrained to be within 0.6 arcseconds of the initial input in both the $x$ and $y$ direction.\\
    $^\mathrm{c}$ The PS magnitude is constrained to be within 7.5 of the initial PS magnitude input.\\
    $^\mathrm{d}$ The position of the PS component is taken to be within 0.15 arcseconds of the host galaxy component (i.e., near the center of the galaxy) in the S\'ersic+PS fit, but is constrained to be within 0.6 arcseconds of the initial input in the PS fit (i.e., treated similar to the host galaxy in the $^\mathrm{b}$ case).}
\end{deluxetable*}

The initial conditions were selected following the work of \citet{Simmons2008}, \citet{Gabor2009}, and \citet{Haussler2011}. We use Source Extractor \citep{Bertin1996} on each image and select the brightest object within a box of side length $1.2"$ at the center of the image, as the cutout is roughly centered on the source to be fit. The source must have an area of 3 pixels each with flux greater than $3\sigma$ over the background as an initial baseline assurance that the source can be fit. From this selection, we find initial values for the position on the image, an effective radius, host magnitude, axis ratio, and position angle. Note that Source Extractor reports different definitions of the axis ratio and position angle than GALFIT. Source Extractor reports the elongation of the source, defined as $a/b$, and thus we take the inverse to be our initial input into GALFIT. The position angle is defined with the positive $x$-axis as $0^{\circ}$ rotating counter-clockwise. Thus we simply shift this value by $90^{\circ}$ to achieve our initial GALFIT input.

We choose a value of $n=2.5$ as the initial S\'ersic index value for all fits. This value is a reasonable midpoint between the historical spiral and elliptical values of $n=1$ and $n=4$, respectively. This value was also selected by \citet{Simmons2008} who studied the viability of the Sérsic+PS fitting method. The initial value for the PS magnitude is taken to be 2 magnitudes dimmer than that of the host galaxy. This method is similar to that of \citet{Gabor2009}. The work of \citet{Haussler2007} heavily tested the robustness of GALFIT on simulated galaxies and found that it is not sensitive to the choice of initial conditions and that the underlying solution is recovered in most cases.

In order to account for nearby sources possibly interfering with the fit, we apply a pixel mask to all pixels that Source Extractor associates with a neighboring source centered outside of $5r_e$ of the primary source. For sources detected within $5r_e$, we perform simultaneous fits including a separate, single S\'ersic component for each additional source. This method adequately minimizes contamination from nearby sources while not prohibitively increasing computing time for a typical cutout.

\begin{figure}[t]
\gridline{\fig{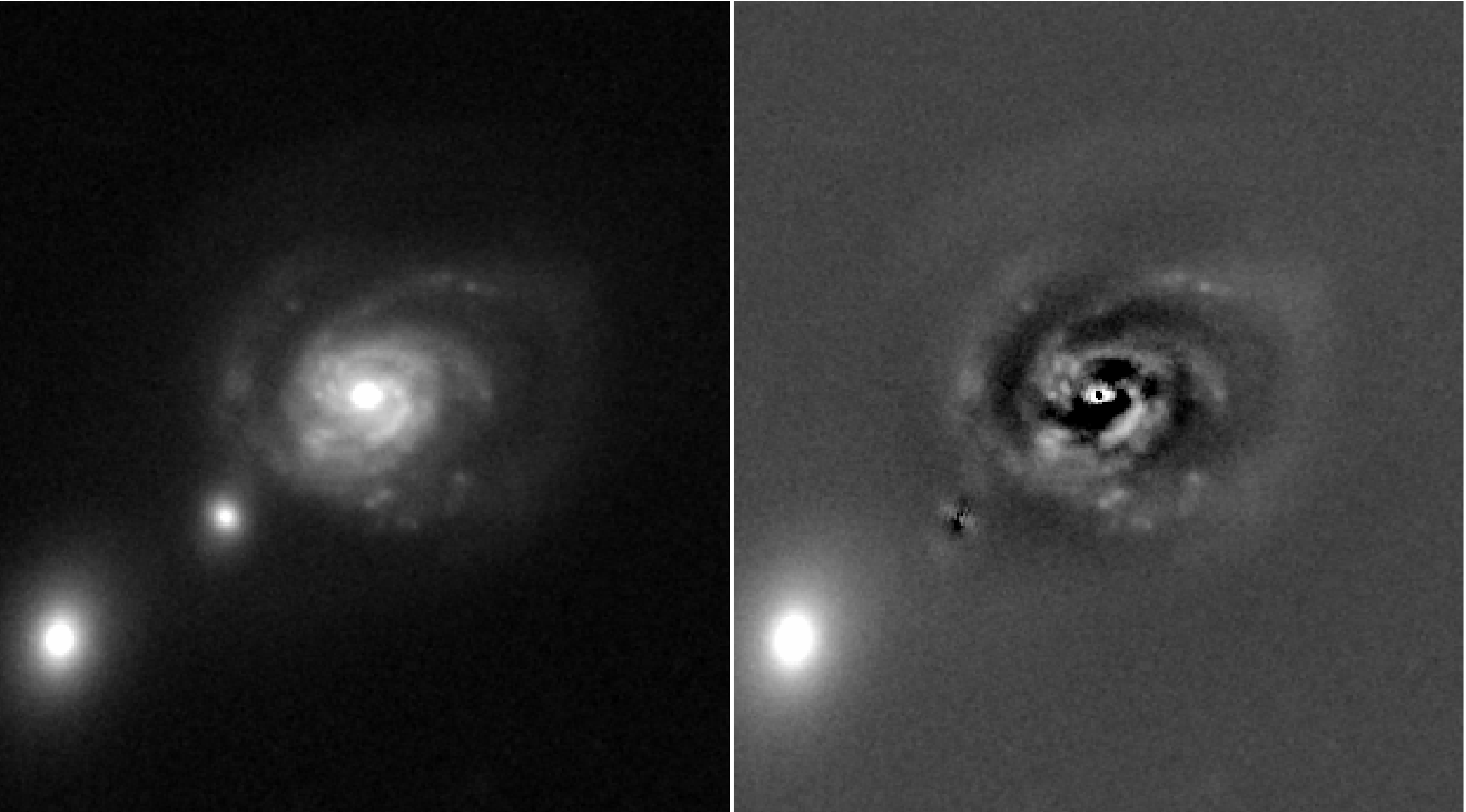}{0.47\textwidth}{Spiral Arms (cid\_3718, 8.91{\arcsec})}}
\gridline{\fig{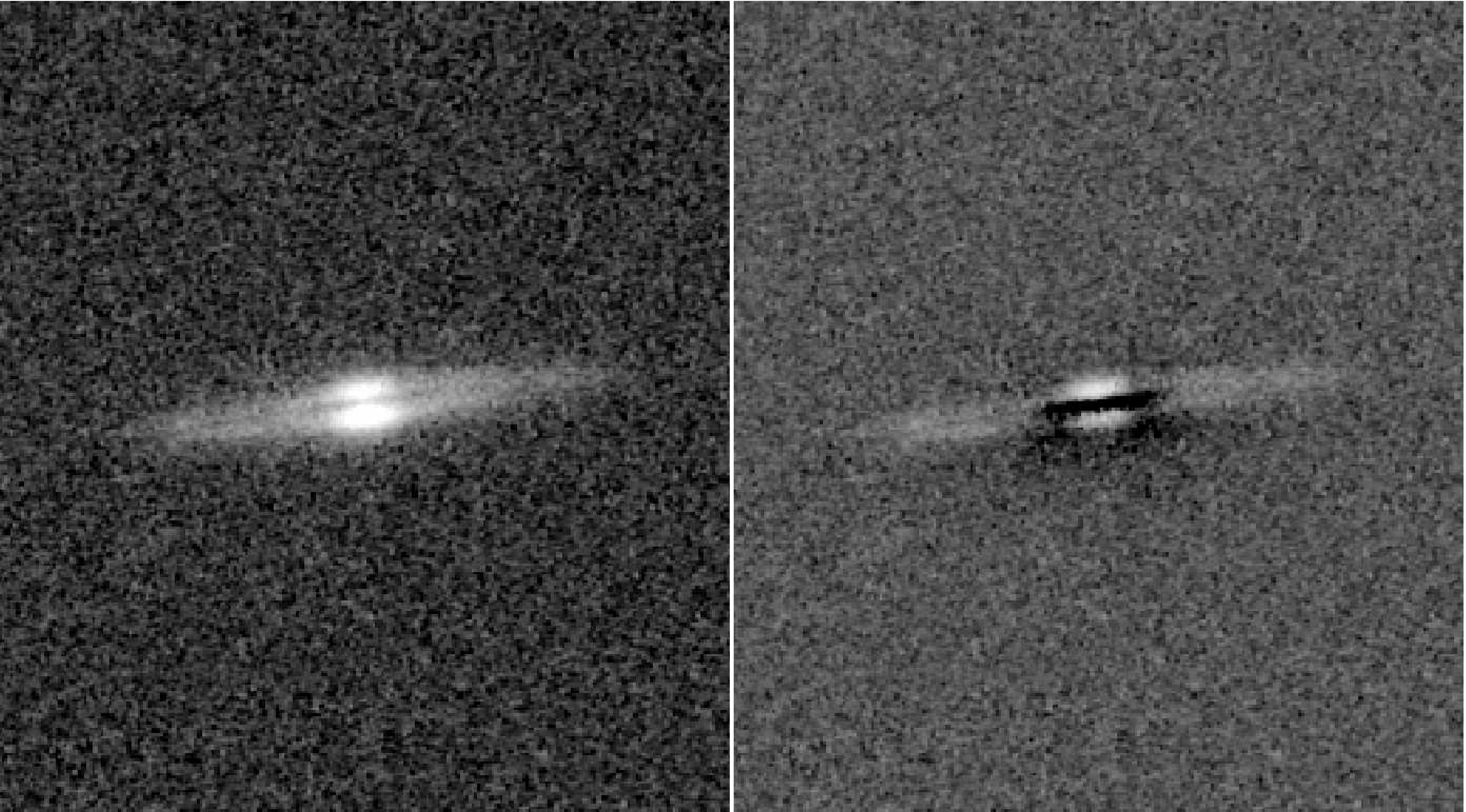}{0.47\textwidth}{Dust Feature (cid\_3439, 7.38{\arcsec})}}
\gridline{\fig{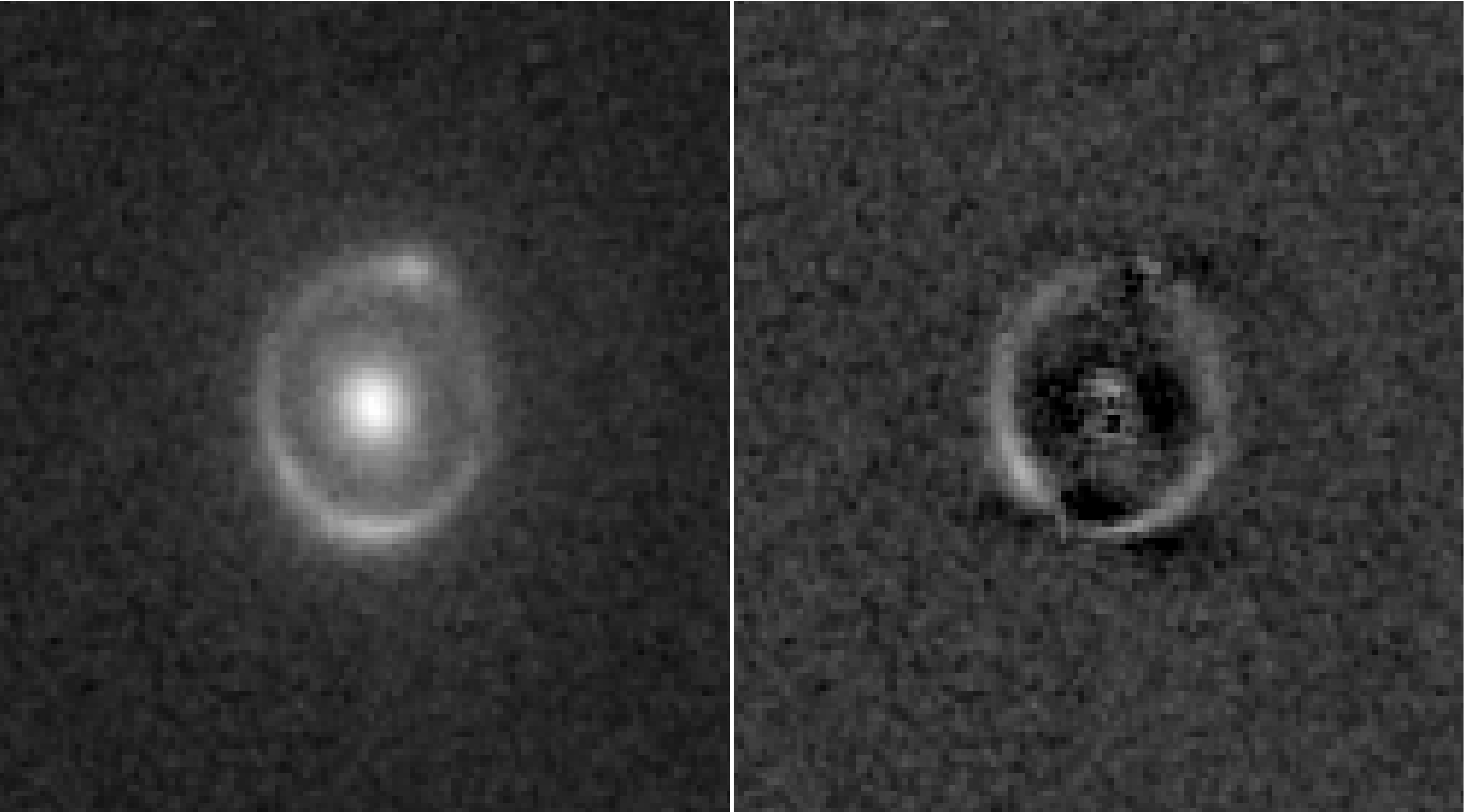}{0.47\textwidth}{Ring (lid\_2072, 4.29{\arcsec})}}
\caption{Examples of features which are unaccounted for in these fits. The cutout (left) is shown alongside the residual of the S\'ersic+PS fit (right). The source ID from the source catalog \citep{Marchesi2016} is given for each example alongside the cutout width.}
\label{fig:example_features}
\end{figure}

The GALFIT Sérsic+PS fitting method cannot account for certain distinctive features in galaxies, such as spiral arms, dust features, or rings. Figure \ref{fig:example_features} shows examples of these alongside an attempted fit. While in many cases GALFIT may return reasonable values for the morphological parameters, these features can influence the fitting algorithm and prevent convergence to physically accepted values.

\section{Testing the Fitting Process} \label{sec:testing}

In order to test our fitting process, we ran a set of fits using the same set of AGN as \citet{Gabor2009} and directly compared the results for each source. This set of data was selected as it uses the same \textit{HST} imaging of the COSMOS field but covers fewer sources, allowing for easy exploration of the quality of fit while also providing the best analog on which to test. The AGN used are optical/IR counterparts from the \textit{XMM-Newton} X-ray source catalog \citep{Cappelluti2007, Brusa2007} and the Very Large Array (VLA) radio source catalog \citep{Schinnerer2007}. This results in a set of 394 AGN which we feed into the fitting process.

\citet{Gabor2009} used similar parameter constraints to the present study. They constrained the PS magnitude to be within 5 magnitudes of the host and to be within 10 magnitudes of the initial value. The radius is also constrained to be less than 500 pixels. The S\'ersic index and PS position constraints are the same between both studies (Table \ref{tab:constraints}). They list no constraints on the host magnitude or position. \citet{Gabor2009} investigated the effects of changing the parameter constraints and found only minor differences in results, validating the viability of this comparison.

Of their 394 fits, \citet{Gabor2009} find that 174 (44\%) of the fits did not converge within the parameter constraints on the initial run. Of these, 74 were re-fit successfully using new initial parameters, leaving only 25\% failing to converge within the parameter constraints. Another 26 fits were later fit manually. Manual fitting refers to creating the input parameters manually without the use of an automated process (e.g., the Source Extractor step). After fitting the set of data using our process, 112 of 394 (28\%) fits failed to converge within the parameter constraints without any re-fitting. Our S\'ersic constraint was flagged 44 times, the maximum radius was flagged 7 times, and the PS position constraints were flagged 36 and 39 times for $x$ and $y$, respectively. In addition, 9 of our fits failed to complete.

\begin{figure*}[t]
\centering
\includegraphics[width=\textwidth]{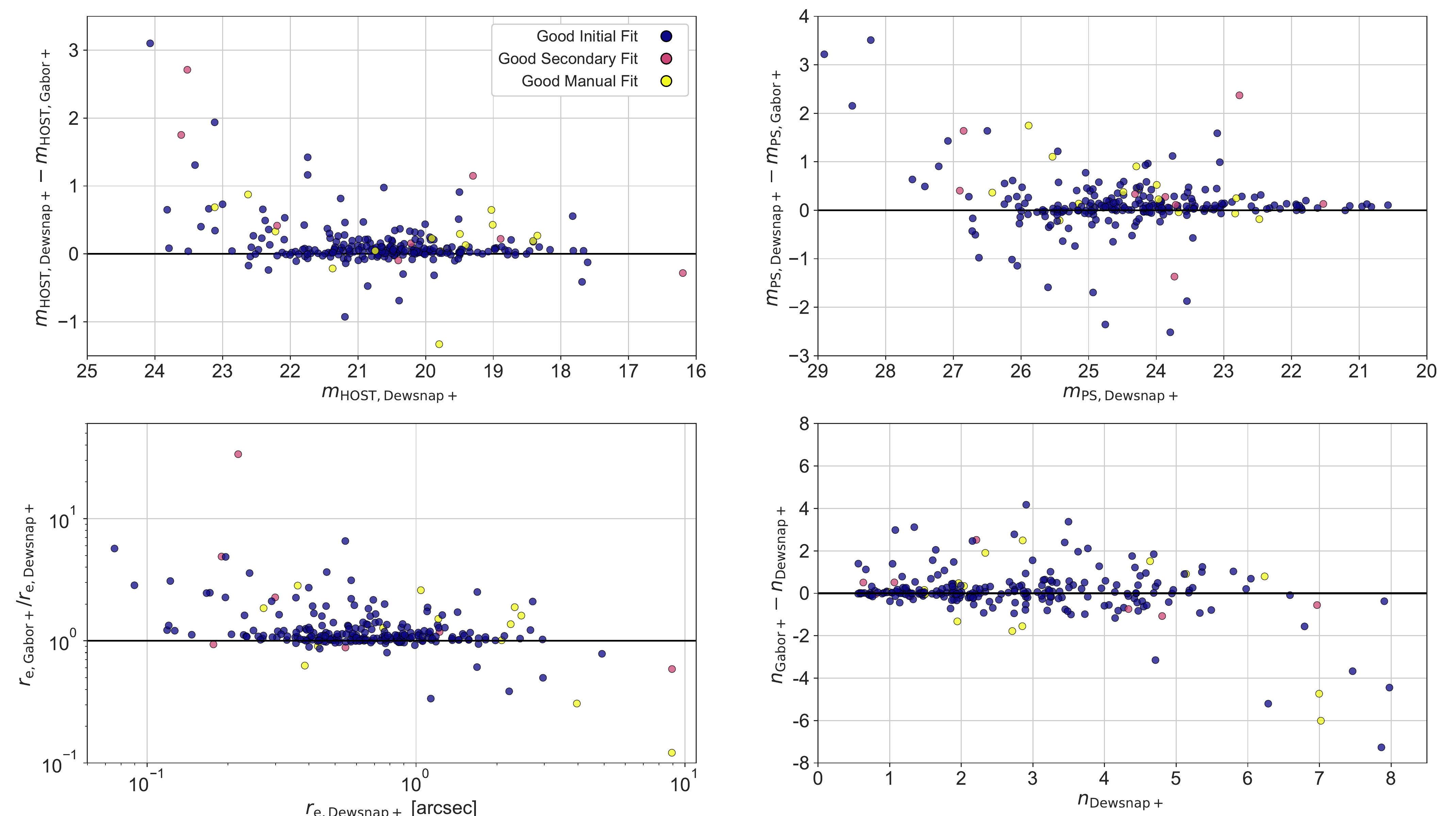}
\caption{Comparison of the test fits and the fits of \citet{Gabor2009}. The sources displayed here are those for which the fits in this study had parameters converge within the constraint boundaries and the fits by \citet{Gabor2009} are flagged as being ``good''. The color of the markers are the fit flag assigned by \citet{Gabor2009}, namely a good initial fit, a good fit after changing the initial parameters, or a good manual fit. The black line represents one-to-one agreement between the fit parameters. All four parameters see strong agreement.}
\label{fig:Gabor_Comparison}
\end{figure*}

Figure \ref{fig:Gabor_Comparison} shows a comparison of the fits from both this study and \citet{Gabor2009}. The fits included are those which did not flag our constraints and those flagged as a good fit by \citet{Gabor2009}. There is clear agreement between the results, with our results being within the uncertainties reported for the majority of fits. The S\'ersic index has the poorest correlation, especially at higher indices. Difficulty in constraining large S\'ersic index values is not uncommon \citep{Ishino2020, Li2021}. Typically, high S\'ersic indices are sensitive to change when the extended wings of the galaxy are faint relative to the background noise \citep{Peng2010}. This may be part of the discrepancy seen here. Overall, these results imply that our fitting process is able to consistently determine best fit parameters which agree with the work of \citet{Gabor2009} for \textit{HST} sources.

\section{Convolved \textit{HST} Fits}

In order to investigate whether our fitting process sees similar results to Subaru with another low resolution set of data comparable to Subaru, we manipulate the \textit{HST} imaging to create a new set of low resolution data that acts as a middle ground between \textit{HST} and Subaru. With this data, we can investigate whether the new convolved \textit{HST} fits are more similar to the \textit{HST} or Subaru fits. If the new convolved fits were to appear more similar to the high resolution \textit{HST} fits, it may unlock new methods to investigate the \textit{HST} and Subaru comparison.

\begin{figure*}[t]
\centering
\includegraphics[width=\textwidth]{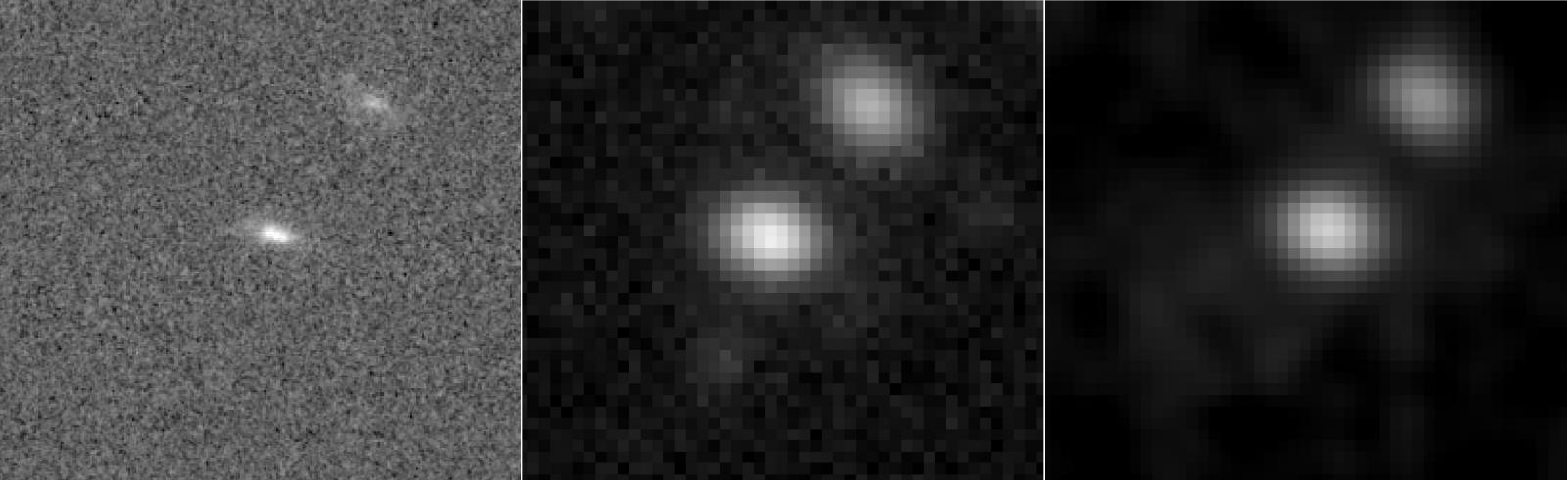}
\caption{Example of the \textit{HST} (left), Subaru (middle), and convolved \textit{HST} (right) cutouts for a typical galaxy, in this case source cid\_1021. Each cutout has a size of 7.65{\arcsec}. Note that the \textit{HST} cutouts have $0.030"/\mathrm{pixel}$ and both the Subaru and convolved \textit{HST} cutouts have $0.168"/\mathrm{pixel}$.}
\label{fig:conv_example}
\end{figure*}

To create the new set of data, we first take the corresponding Subaru PSF and use Montage \citep{Jacob2010} to scale the pixel scale to match that of the \textit{HST} cutouts. Then, we convolve the \textit{HST} cutout with the scaled Subaru PSF using Astropy \citep{AstropyCollaboration2013, AstropyCollaboration2018}. The final step is to scale the new image to match the pixel scale of the Subaru cutouts. In addition, the image flux must be scaled by a factor equal to the ratio of the new and old pixel area, namely $\left(\frac{0.168}{0.030}\right)^2$. Figure \ref{fig:conv_example} demonstrates a comparison of a source imaged by both \textit{HST}, Subaru, and our new convolved image. Visually, the convolved \textit{HST} image appears more similar to the Subaru cutout.

As these images are a combination of \textit{HST} and Subaru imaging, the PSF is also a combination of the \textit{HST} and Subaru PSFs. We created a new convolved \textit{HST} PSF following a similar process to creating the cutouts. We first adjust the pixel scale of the corresponding Subaru PSF to match that of \textit{HST}, then convolve the \textit{HST} PSF with the scaled Subaru PSF. We then adjust the scale of the new PSF to match that of Subaru. These are the PSFs we used with GALFIT in order to fit these cutouts.

\section{Results} \label{sec:results}

We applied our fitting process to each of the \textit{HST}, Subaru, and convolved \textit{HST} cutouts. The time required for each source depends on the number of steps taken to converge and the size of the cutout. A source with more nearby sources will, in general, take more steps to converge, as it may take longer to fit the neighboring sources than the original source. The angular size of the cutout is a factor as well, as a larger angular size likely includes more neighboring sources and thus requires more steps. A cutout with a larger number of pixels will also increase computation time as, in general, a larger cutout contains a larger source. Thus, a larger image will require more convolutions per iteration, which is the most time consuming step in the fitting process. In general, our Subaru cutouts cover a larger angular region, meaning that there are typically more nearby sources to fit. Our \textit{HST} sources, however, tend to have a larger number of pixels. This leads to smaller \textit{HST} cutouts taking the least time to fit (on the order of seconds), followed by most Subaru cutouts (seconds to minutes), with larger \textit{HST} cutouts taking the longest (minutes to hours). In general, the convolved \textit{HST} fits were always the quickest to compute, as they feature the smaller angular region of the \textit{HST} cutouts but also the lower pixel scale of the Subaru cutouts.

Of the 4016 sources in the catalog, 2782 and 2995 sources passed the Source Extraction step defined in Section \ref{sec:process} for \textit{HST} and Subaru, respectively. This step does not complete for sources that either do not have a sufficiently bright \textit{i}-band counterpart or are not near the center of their respective cutout. Of the 2782 \textit{HST} sources passed to GALFIT, 66 S\'ersic+PS fits failed to converge, alongside 71 and 80 for the single S\'ersic and PS fits, respectively. For the 2988 Subaru sources passed to GALFIT, there were 138 failed S\'ersic+PS fits, and 112 and 145 failed single S\'ersic and PS fits, respectively. It is not unexpected for the Subaru fits to fail marginally more often than the \textit{HST} fits due to the reduced spatial resolution. In comparison, the convolved \textit{HST} fits had 2803 sources pass the Source Extraction step, with an additional 85 S\'ersic+PS fits failing to converge.

\subsection{Parameter Constraints}

\begin{deluxetable*}{lDDD}
\tablecaption{The distribution of constraints flagged by each S\'ersic+PS fit. \label{tab:constraints_hit}}
\tablehead{
\colhead{} & \multicolumn{6}{c}{Number (\%) of Constraints Flagged}
\\
\colhead{Parameter} & \multicolumn{2}{c}{\textit{HST} (2716 Fits)} & \multicolumn{2}{c}{Subaru (2857 Fits)} & \multicolumn{2}{c}{Convolved \textit{HST} (2718 Fits)}}
\decimals
\startdata
    None & 1871 \ (68.89\%) & 1377 \ (48.20\%) & 1359 \ (50.00\%) \\
    S\'ersic Index & 640 \ (23.56\%) & 806 \ (28.21\%) & 678 \ (24.94\%) \\
    Host Magnitude & 76 \ (2.80\%) & 96 \ (3.36\%) & 18 \ (0.66\%) \\
    Half-Light Radius & 25 \ (0.92\%) & 170 \ (5.95\%) & 5 \ (0.18\%) \\
    $x_\mathrm{host}$ Position & 7 \ (0.26\%) & 39 \ (1.37\%) & 15 \ (0.55\%) \\
    $y_\mathrm{host}$ Position & 5 \ (0.18\%) & 37 \ (1.30\%) & 20 \ (0.74\%) \\
    Point-Source Magnitude & 73 \ (2.69\%) & 59 \ (2.07\%) & 6 \ (0.22\%) \\
    $x_\mathrm{PS}$ Position & 163 \ (6.00\%) & 804 \ (28.14\%) & 913 \ (33.59\%) \\
    $y_\mathrm{PS}$ Position & 151 \ (5.56\%) & 826 \ (28.91\%) & 949 \ (34.92\%) \\
\enddata
\tablecomments{A fit can flag multiple constraints (e.g., a fit can flag both the S\'ersic index and $x_\mathrm{PS}$ position constraint flags. Both flags in the single are counted in this table -- thus the columns sum to a value $\geq100\%$). See Table \ref{tab:constraints} for the definition of each constraint.}
\end{deluxetable*}

Of these successful fits, 31\% of \textit{HST} S\'ersic+PS fits failed to converge within the parameter constraints compared to 52\% for Subaru. The convolved \textit{HST} S\'ersic+PS fits failed to converge within the bounds in 50\% of the fits, a value very similar to that of Subaru. Table \ref{tab:constraints_hit} gives a summary of which specific constraints are flagged for each of these fits. For \textit{HST}, the S\'ersic index did not fall within the constraints more often than any other parameter. This result is not entirely unexpected, as the S\'ersic index tends to increase to high values in cases where the source appears point-like, thus often running into the upper constraint. The upper limit of the S\'ersic index is also frequently reached in the Subaru fits, however the PS position is far more likely to reach the constraint boundaries in the Subaru fit compared to the \textit{HST} fit. This is likely due to the smaller number of pixels (although same angular distance) that the PS component is free to move in the Subaru fits compared to the \textit{HST} fits. Also, it may be harder to separate where the AGN lies in the bulge, as the bulge and AGN can appear to be of similar size due to the broader PSF of Subaru. The upper radius constraint is also far more likely to be flagged for Subaru. This is seen in many point-like galaxies with high S\'ersic indices. This is because as the S\'ersic index increases, the more point-like the profile becomes, even with large effective radii. Low values for the S\'ersic index are also seen in these point-like galaxies -- in some cases the S\'ersic index minimizes with this large radii and forms a flatter, fainter model allowing the PS component to essentially model both the inner galaxy and the AGN. The convolved \textit{HST} S\'ersic+PS fits appear to flag a similar number of constraints as the \textit{HST} and Subaru fits. The majority of the convolved \textit{HST} constraints are more comparable to the \textit{HST} results, however the PS component position constraints are more comparable to Subaru.

For the single S\'ersic component fits, we find that 31\% of the \textit{HST} fits and 33\% of the Subaru fits flagged the parameter constraints. These are slightly lower values than those seen in the S\'ersic+PS fits. This is not unexpected, as there are fewer constraints that are possible to be flagged. Note that a higher number of these fits flag the S\'ersic constraint, but there are still fewer constraints flagged in total due to the lack of a point-source component. For the single PS fits, we find that 4\% of the \textit{HST} fits and 13\% of the Subaru fits flagged the constraints. Again, this is not unexpected. The parameter constraints for the single PS fit roughly correspond to the host galaxy position and magnitude constraints seen in the S\'ersic+PS fits which are rarely flagged.

\subsection{Morphological Parameters} \label{subsec:parameters}

In order to determine the quality of the fits, we apply a set of cuts based on the results. This is to both establish whether the source itself is expected to provide a reliable fit (e.g., we don't expect a very faint source to provide a reliable fit) and if the morphological parameters output by GALFIT are physically meaningful. The cuts are based on those used in similar studies \citep{Simmons2008, Gabor2009}.

The first cuts are based on a subset of the parameter constraints. If a fit has any of the magnitude, effective radius, or S\'ersic index constraint flags then it is cut from the set of ``good'' fits. The position constraint flags are excluded from the cuts as they don't necessarily relate to quality of fit; their primary function is simply to ensure that the correct source is being fit. The rest of the cuts are based on the output morphological parameters. First, sources with $m_\mathrm{HOST} > 25$ or $m_\mathrm{PS} > 28$ are excluded. Additionally, fits with $r_\mathrm{e} < 0.015"$ are also cut. We do not apply any cuts based on the $\chi^2_\nu$ of the fit as we find that it does not adequately probe for poor fits compared to the other parameter cutoffs. A wide range of $\chi^2_\nu$ limits were tested and none saw consistent improvement on the quality of the ``good'' fits.

Of the 2716 successful S\'ersic+PS \textit{HST} fits, 427 (15.72\%) are cut for $r_\mathrm{e} < 0.015"$, 310 (11.41\%) for $m_\mathrm{HOST} > 25$, and 453 (16.68\%) for $m_\mathrm{PS} > 28$. In combination with the constraint-based cuts (see Table \ref{tab:constraints_hit}), 1367 (50\%) of the \textit{HST} S\'ersic+PS fits are cut. Comparatively, the 2857 Subaru S\'ersic+PS fits have 493 (17.26\%), 241 (8.44\%), and 157 (5.50\%) fits cut for the radius, host magnitude, and PS magnitude limits. In total, 1521 (53\%) Subaru S\'ersic+PS fits are cut.

\begin{figure*}[t]
\centering
\includegraphics[width=\textwidth]{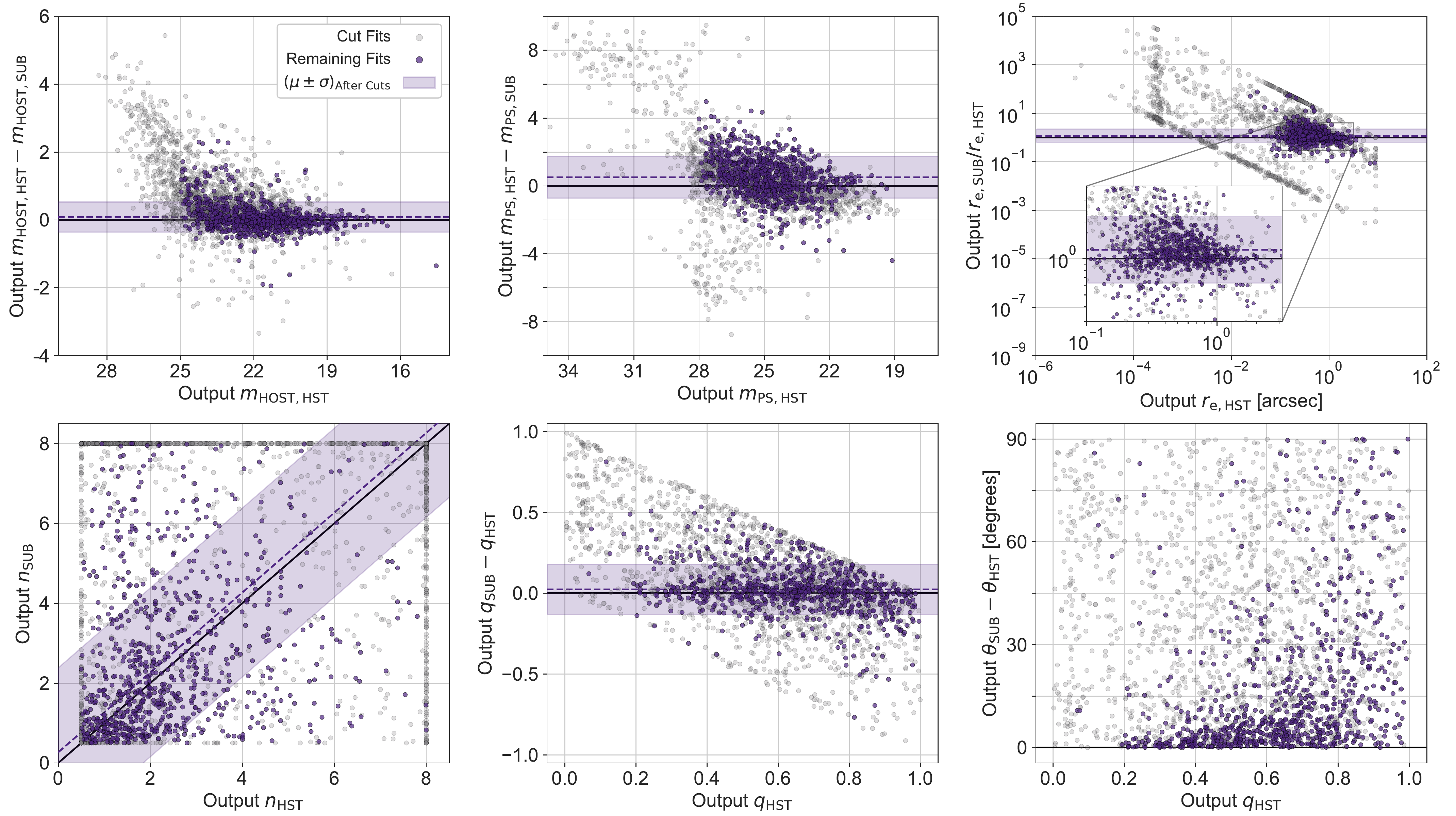}
\caption{Comparison of the \textit{HST} and Subaru S\'ersic+PS fits. The points marked in gray include all fits for both \textit{HST} and Subaru, regardless of whether the fit failed to pass the parameter/constraint cuts. The points marked in purple are those which are flagged as ``good'' for both \textit{HST} and Subaru. The solid, black line in each plot represents agreement between the fits. The dashed, purple line and associated shaded region represent the average distance from agreement. Upper-left: difference in output host magnitude measurement between \textit{HST} and Subaru vs. the output \textit{HST} host magnitude. Upper-middle: difference in output central point-source magnitude measurement between \textit{HST} and Subaru vs. the output \textit{HST} point-source magnitude. Upper-right: ratio of the output effective radii vs. the output \textit{HST} effective radius (in arcsec). Note that the purple line and region are determined in log-space for this case. Lower-left: output Subaru S\'ersic index vs. the output \textit{HST} S\'ersic index. Lower-middle: difference in output axis ratio ($b/a$) between Subaru and \textit{HST} vs. the output \textit{HST} axis ratio. Lower-right: absolute difference in output position angle (in degrees) between Subaru and \textit{HST} vs. the output \textit{HST} axis ratio. Note that this plot does not have the purple line and region; this is because one expects a larger difference in position angle as the axis ratio approaches $q=1$, as the source becomes more circular and thus the position angle becomes increasingly irrelevant. Thus, the region would not be representative of the difference in the position angle measurements.}

\label{fig:out_vs_out_sub}
\end{figure*}

We begin comparing the \textit{HST} and Subaru results by directly comparing the spatial parameters returned by each fit. Figure \ref{fig:out_vs_out_sub} demonstrates this comparison for each of the major parameters returned by GALFIT for the S\'ersic+PS fits. There is clearly less agreement than seen in the testing process (Figure \ref{fig:Gabor_Comparison}), but there is good agreement for many parameters. Note that the testing process involves comparing two sets of \textit{HST} fits, so closer agreement than our \textit{HST}--Subaru comparison is expected.

The host magnitude values are in good agreement for the vast majority of S\'ersic+PS fits. This is true especially for the fits which pass the parameter/constraint cuts, although the majority of the cut fits also see significant agreement. At $m_\mathrm{HOST,HST} \sim 25$ there is a clear break in the agreement; this break helps justify the choice of the host magnitude cutoff point. On average, we find that the output Subaru host magnitudes are brighter than their respective \textit{HST} host magnitudes by only $0.08 \pm 0.45 \ \mathrm{mag}$ for the ``good'' fits.

The PS magnitude measurements also tend to agree, however the distribution is wider. There are primarily two regions of noticeable disagreement: Subaru finds many $m_\mathrm{PS,SUB}$ significantly brighter at $m_\mathrm{PS,HOST} > 28$ and many significantly dimmer between $28 > m_\mathrm{PS,HOST} > 25$. These two groups are caused by the same phenomenon -- they are both regions where one of the \textit{HST} or Subaru S\'ersic+PS fits failed to separate the AGN from its host galaxy. This causes GALFIT to dim the PS component beyond the limiting point-source depth of the images. This occurs most commonly for sources which have no distinct point-source or for point-like galaxies where there is no distinct host galaxy. In many of the cases where the source appears point-like, the S\'ersic index reaches very high values. This is because a S\'ersic profile appears more point-like as $n$ increases. There is a significant offset from one-to-one agreement; brighter $m_\mathrm{PS,HST}$ see close agreement to $m_\mathrm{PS,SUB}$, however as $m_\mathrm{PS,HST}$ becomes dimmer $m_\mathrm{PS,SUB}$ tends to be brighter. Thus, $m_\mathrm{PS,SUB}$ is, on average, brighter than $m_\mathrm{PS,HST}$ by $0.51 \pm 1.24 \ \mathrm{mag}$ for the ``good'' S\'ersic+PS fits.

The output effective radii covers a comparatively large parameter space, very little of which corresponds to agreement between the fits. Nonetheless, the parameter cuts again remove the majority of fits with strong disagreement. Most notable of the cut fits are regions of very low radius. The majority of the cut fits are grouped at $r_\mathrm{e} \sim 0.01$ pixels for either (or both) of \textit{HST} or Subaru. This value appears to be a minimum value allowed by GALFIT. The possible cause of these groups are discussed in Section \ref{sec:discussion}. For the ``good'' fits, we again see significant agreement. Interestingly, we see agreement down to the minimum radius value of $0.015"$, despite this being significantly smaller than a Subaru pixel ($0.168"/\mathrm{pixel}$). On average, the Subaru radius is $1.18 \pm 1.88$ times larger than the \textit{HST} radius. Note that this average was determined in log-space.

\begin{figure*}[t]
\centering
\gridline{\fig{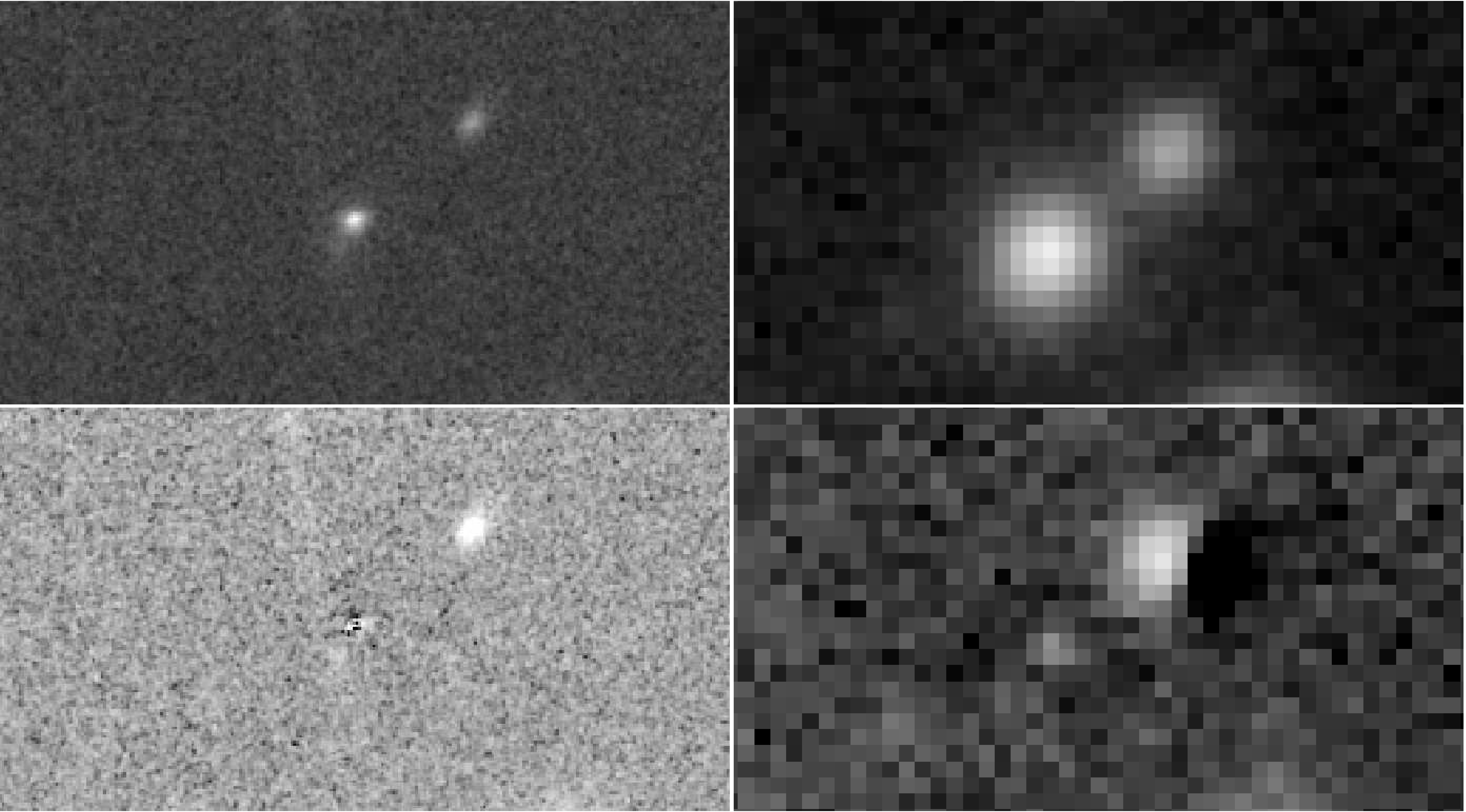}{0.48\textwidth}{(a) cid\_1008, 7.65{\arcsec} \\ \textline[t]{$q_{HST}=0.64$}{$q_\mathrm{SUB}=0.96$} \textline[t]{$\theta_{HST}=141^{\circ}$}{$\theta_\mathrm{SUB}=27^{\circ}$}}
\fig{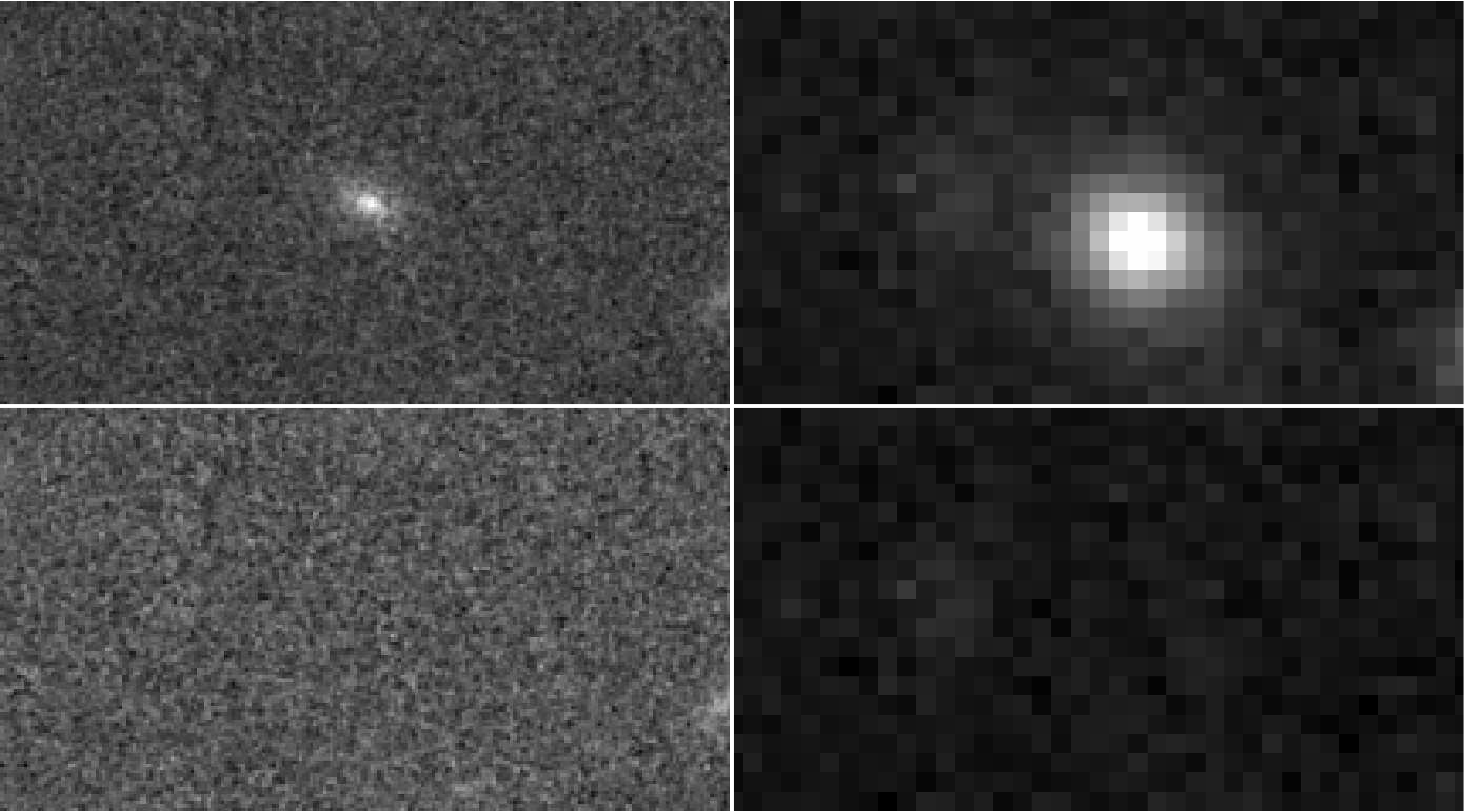}{0.48\textwidth}{(b) cid\_1051, 6.46{\arcsec} \\ \textline[t]{$q_{HST}=0.56$}{$q_\mathrm{SUB}=0.84$} \textline[t]{$\theta_{HST}=63^{\circ}$}{$\theta_\mathrm{SUB}=57^{\circ}$}}}
\caption{Comparison of the axis ratio and position angles (using the GALFIT definitions from Section \ref{sec:process}) output by Source Extractor for two different sources as measured on \textit{HST} and Subaru imaging. The upper images in each grid represents the image cutout, whereas the lower images correspond to the residual of the GALFIT fit. The leftmost images in each grid are the \textit{HST} images and the right are Subaru. The source IDs from \citet{Marchesi2016} are provided for each example alongside the cutout width. Note that the \textit{HST} cutouts have $0.030"/\mathrm{pixel}$ and the Subaru cutouts have $0.168"/\mathrm{pixel}$.}
\label{fig:axis_comp}
\end{figure*}

The axis ratio and position angle measurements are inherently linked, and thus it is impossible to compare the results of each separately. Looking first at the axis ratio, we see that the majority of the fits which were not cut see significant agreement, with an average difference of $q_\mathrm{SUB} - q_\mathrm{HST} = 0.02 \pm 0.15$. Some scatter is expected in this comparison due to the difference in resolution between the two telescopes. A single Subaru pixel is equal in area to ${\sim}$32 \textit{HST} pixels; a galaxy made up of hundreds of \textit{HST} pixels comprises of only a few Subaru pixels, making the axis ratio much more difficult to determine. Since the position angle depends on the measurement of the axis ratio, it also faces similar difficulty. Examples of this can be seen in Figure \ref{fig:axis_comp}. In order to compare the position angle measurement, we must contextualize it with respect to the axis ratio. If $q \sim 1$, then the source appears roughly circular and the position angle is meaningless. If the axis ratio is lower, the difference in position angle is more striking, and thus we expect stronger agreement. Thus, we expect no level of agreement as the axis ratio increases towards this value. In Figure \ref{fig:out_vs_out_sub}, we see this relationship clearly for the ``good'' fits; at lower axis ratios we see stronger agreement whereas at high ratios we see very little agreement.

\begin{figure*}[t]
\centering
\includegraphics[width=\textwidth]{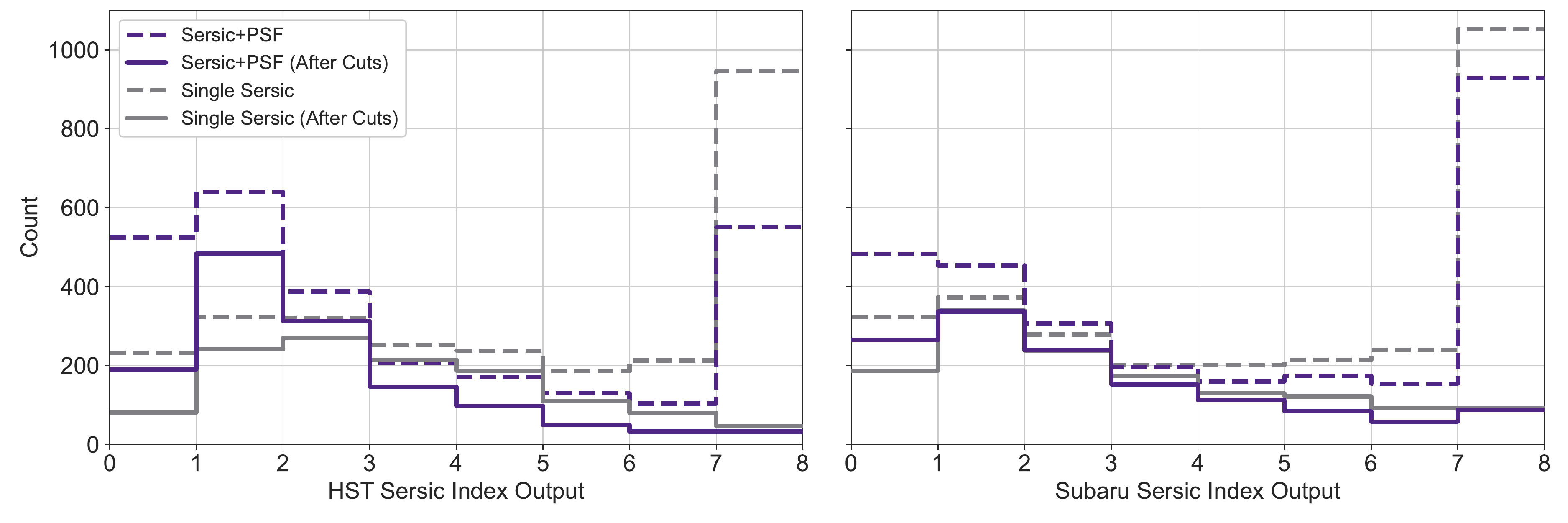}
\caption{Comparison of the S\'ersic indices measured by the single S\'ersic component and the S\'ersic+PS fits for both \textit{HST} (left) and Subaru (right). The purple lines represent the two component S\'ersic+PS fits and the gray represents the single component S\'ersic fits. The dashed lines represents all fits which completed successfully, while the solid lines only include fits which passed the parameter/constraint cuts for \textit{HST} fits (left) or Subaru fits (right).}
\label{fig:sersic_hist}
\end{figure*}

The parameter which disagrees the most between the two sets of fits is the S\'ersic index. There is essentially no agreement between the \textit{HST} and Subaru fit results. There does exist a group of similarly low-valued fits between $0.5<n<3$ populated primarily by the ``good'' fits. The possible cause of this surprising result is discussed in more depth in Section \ref{sec:discussion}. Figure \ref{fig:sersic_hist} demonstrates the distribution of the S\'ersic indices for each set of fits. The vast majority of fits which failed the parameter/constraint cuts are at the $n=8$ limit, especially in the single S\'ersic component fits. In the case of \textit{HST}, the single-component fits tend to give a higher S\'ersic index than the corresponding two-component fits. This is an expected result, as by not accounting for the central point-source, more flux is incorrectly attributed to the bulge of the galaxy, thus resulting in a higher S\'ersic index \citep{Simmons2008}. We see a similar result in the Subaru fits, although to a much lower degree.

\begin{figure*}[t]
\centering
\includegraphics[width=\textwidth]{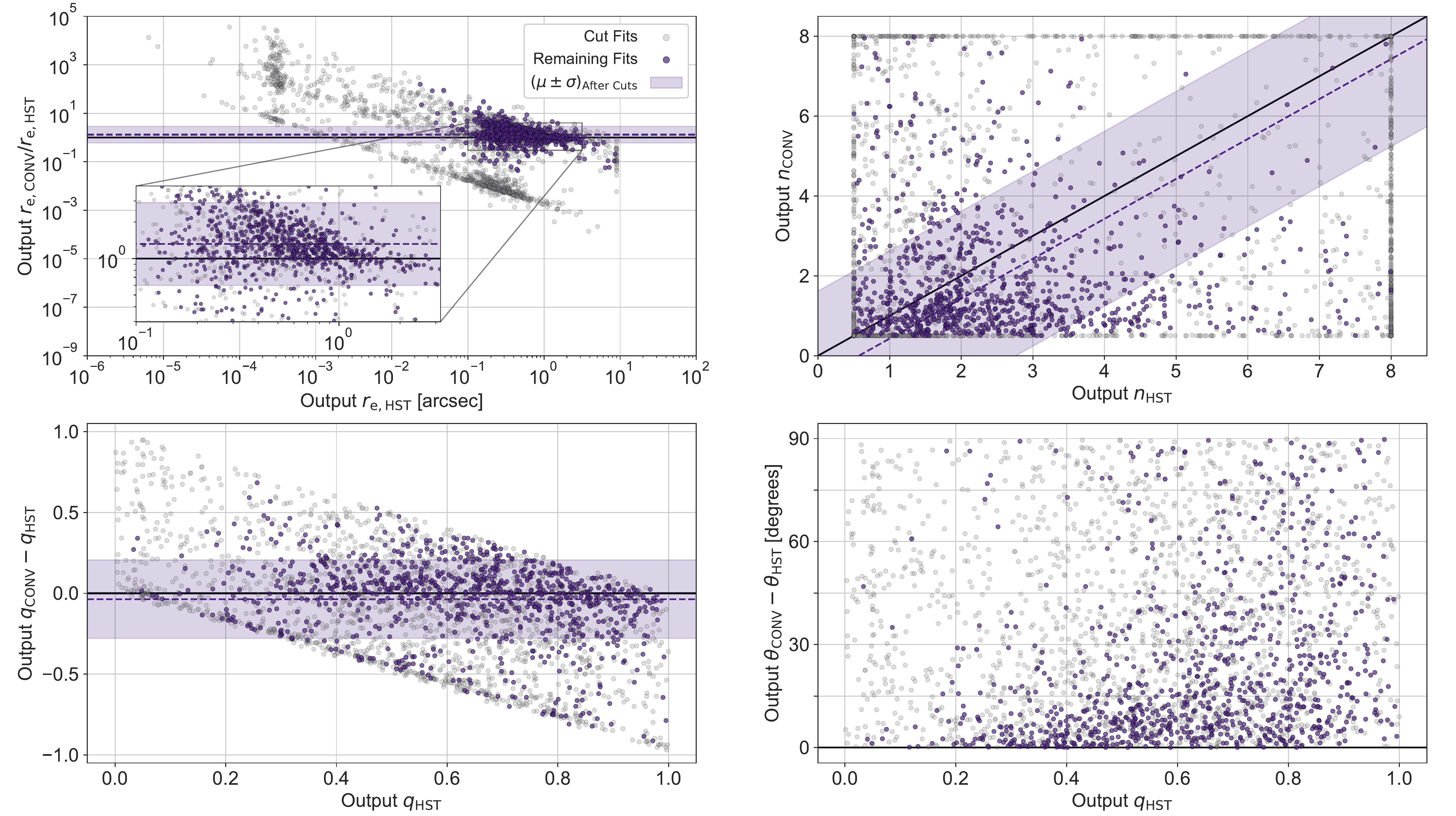}
\caption{Comparison of the \textit{HST} and convolved \textit{HST} S\'ersic+PS fits. The points marked in gray include all fits for both \textit{HST} and convolved \textit{HST}, regardless of whether the fit failed to pass the parameter/constraint cuts. The points marked in purple are those which are flagged as ``good'' for both \textit{HST} and Subaru. See Figure \ref{fig:out_vs_out_sub} for an explanation of the respective plots.}
\label{fig:out_vs_out_conv}
\end{figure*}

Figure \ref{fig:out_vs_out_conv} shows the direct comparison between the output parameters of the \textit{HST} and convolved \textit{HST} fits. We see a very similar result to that of Figure \ref{fig:out_vs_out_sub}. As seen in Table \ref{tab:constraints_hit}, far fewer fits reach the upper radius constraint. However, we see far more fits which converge to the lower radius limit seen at 0.01 pixels, as previously discussed for the \textit{HST}--Subaru comparison above (Figure \ref{fig:out_vs_out_sub}). In addition, far more convolved \textit{HST} fits reach extremely low axis ratios $(q_\mathrm{CONV}\lesssim0.1)$ than seen for either \textit{HST} or Subaru, including many which are classified as ``good'' fits.

Most notably, Figure \ref{fig:out_vs_out_conv} shows virtually no agreement in the S\'ersic index, with the disagreement appearing very similar in form to the \textit{HST}--Subaru S\'ersic index comparison. Importantly, we find that there are similar levels of agreement between the convolved \textit{HST} and Subaru fits as between \textit{HST} and Subaru. Notably, the fits which disagree between Subaru and \textit{HST} in terms of host magnitude and radius tend to also disagree between Subaru and the convolved \textit{HST}. There is also no agreement between the Subaru and convolved \textit{HST} S\'ersic index. This implies that the issue in determining the S\'ersic index is likely due to the broader PSF seen in both low resolution sets of data.

\subsection{Multi-Component vs. Single Component Fits}

\begin{figure*}[t]
\centering
\includegraphics[width=\textwidth]{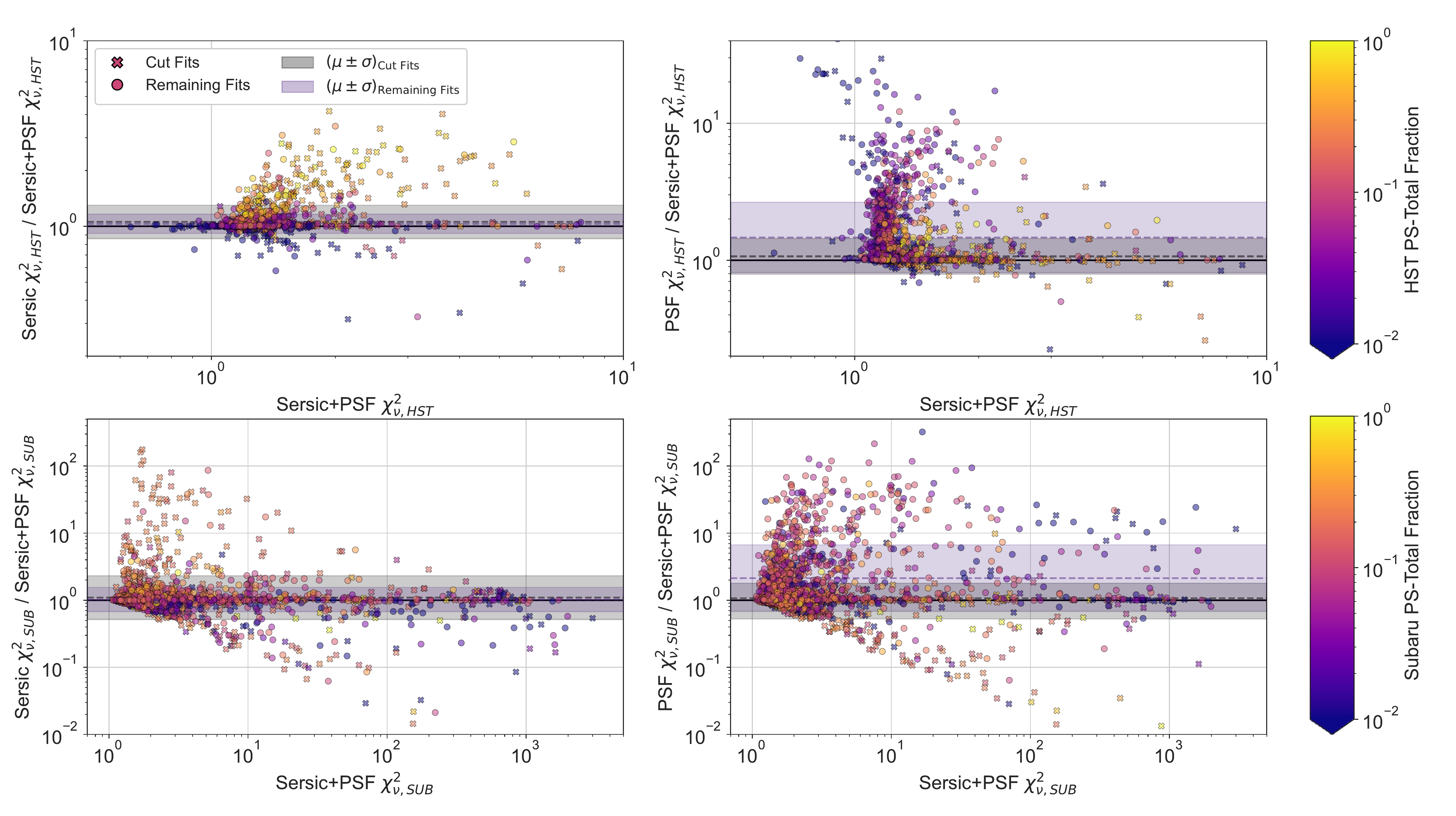}
\caption{Comparison of the reported $\chi^2_\nu$ between the three fits for both \textit{HST} and Subaru. The upper plots are for \textit{HST}, and the lower for Subaru. The left plots show the difference in $\chi^2_\nu$ between the S\'ersic+PS and single S\'ersic fits, and the right plots show the difference between the S\'ersic+PS and the single PS fits. The colour of the markers represent the fraction of flux from the point-source relative to the combined flux of the point-source and the host galaxy (i.e., yellow represents point-source dominated, purple represents galaxy dominated). The solid, dashed line represents one-to-one agreement between the fits. The circular markers represent the fits which were marked as ``good'' for their respective telescope's S\'ersic+PS fit, whereas the crossed markers represent those which did not pass the set of cuts. The shaded regions represent the average difference between the $\chi^2_\nu$ of the corresponding fits, with the purple and gray regions representing the fits which did and did not pass the parameter/constraint cuts, respectively. Here we see the expected relation between $\chi^2_\nu$ and fit type for \textit{HST}, where the single S\'ersic fits have higher $\chi^2_\nu$ than the corresponding S\'ersic+PS fit. Likewise, we find that the single PS fits yield higher $\chi^2_\nu$ for galaxy-dominated sources. We see no such strong relation for the Subaru fits.}
\label{fig:chi2nu_results_sub}
\end{figure*}

By computing the three different types of fits, as described in Section \ref{sec:process}, we are able to determine whether either of the two single-component fits (the single S\'ersic or single PS fit) performed better than the two-component S\'ersic+PS fits, using the $\chi^2_\nu$ for the respective fits. By the definition of $\chi^2_\nu$ given in Equation \ref{eq:chi2nu}, we see that a good fit should have a $\chi^2_\nu$ value near 1. In general, a value greater than 1 implies that the model is not fitting the source well, and a value less than 1 implies over-fitting of the noise. However, in the case of the \textit{HST} and, by extension, the convolved \textit{HST} imaging, this is not entirely true. The definition of $N_\mathrm{dof}$ in Equation \ref{eq:chi2nu} assumes that all of the pixels in the image are independent. Since the \textit{HST} ACS/WFC data is drizzled from $0.05"/\mathrm{pixel}$ to $0.03"/\mathrm{pixel}$, not all of the pixels are independent. For these fits, $N_\mathrm{dof}$ must be redefined using the number of pixels contained in the raw image rather than the drizzled image. Thus, the reported $\chi^2_\nu$ from GALFIT will have ideal fits with $\chi^2_\nu < 1$. The reported $\chi^2_\nu$ is defined as
\begin{equation}
    \chi^2_\nu = \frac{\chi^2}{N_\mathrm{dof}} = \frac{\chi^2}{n_x n_y - N_\mathrm{mask} - N_\mathrm{\alpha}},
\end{equation}
where $\chi^2_\nu$ is the reduced $\chi^2$ reported by GALFIT, $n_x$ and $n_y$ are the number of pixels in the $x$ and $y$ directions, $N_\mathrm{mask}$ is the number of pixels removed by the mask, and $N_\mathrm{\alpha}$ is the number of free parameters. The true reduced $\chi^2$ is defined as
\begin{equation}
   \overline{\chi^2_\nu} = \frac{\chi^2}{\left(n_x n_y - N_\mathrm{mask}\right)  \left( \frac{0.03}{0.05} \right)^2 - N_\mathrm{\alpha}},
\end{equation}
where the additional $\left( \frac{0.03}{0.05} \right)^2$ term converts from the number of drizzled pixels to the number of raw pixels. Using the common $\chi^2$ between these expressions, we can use the reported $\chi^2_\nu$ to determine the true $\overline{\chi^2_\nu}$ using
\begin{equation}
    \overline{\chi^2_\nu} = \frac{n_x n_y - N_\mathrm{mask} - N_\mathrm{\alpha}}{\left(n_x n_y - N_\mathrm{mask}\right)  \left( \frac{0.03}{0.05} \right)^2 - N_\mathrm{\alpha}} \chi^2_\nu.
\end{equation}
An ideal \textit{HST} fit will have $\overline{\chi^2_\nu} = 1$. Since, in general, $\left(n_x n_y - N_\mathrm{mask}\right) \gg N_\mathrm{\alpha}$, we can approximate that an ideal fit will have a reported $\chi^2_\nu = \left( \frac{0.03}{0.05} \right)^2 = 0.36$. Throughout this work, we primarily discuss the reported $\chi^2_\nu$ as that is the value GALFIT directly interacts with.

Figure \ref{fig:chi2nu_results_sub} shows the comparison between the reported $\chi^2_\nu$ for each of the fit types for each of \textit{HST} and Subaru. The \textit{HST} plots show that there are very few fits for which the S\'ersic+PS fit does not have a lower value for $\chi^2_\nu$ than the single component fits. The single S\'ersic fit performs marginally better for a small number $\left({\sim}2\%\right)$ of fits for which the point-source is faint relative to the host galaxy. There exists a larger subset of fits $\left({\sim}15\%\right)$ for which the single S\'ersic profile performs notably worse than than the S\'ersic+PS, all of which are point-source dominated. Looking at the single PS \textit{HST} fits, there are only a small number of fits with a lower $\chi^2_\nu$ than their respective S\'ersic+PS fit, again ${\sim}2\%$. These sources tend to be more point-source dominated. There does exist another large group $\left({\sim}25\%\right)$ of fits for which the single PS fit performs much worse than the S\'ersic+PS fits, most of which have a low point-source fraction. These \textit{HST} results are expected and self-explanatory; a point-source-dominated object is fit well by a single PS and a galaxy-dominated source is fit well by a single S\'ersic profile. 

In the Subaru fits, we see no trends of this type. The single-component fits do not follow any trend with the point-source fraction as with the \textit{HST} fits. There is a much larger number of Subaru sources whose single S\'ersic component fit provides a lower $\chi^2_\nu$ than the S\'ersic+PS fit $\left({\sim}8\%\right)$, but this occurs seemingly indiscriminately. There are a similar number of fits whose S\'ersic+PS fit provides a lower $\chi^2_\nu$ than the single S\'ersic fit, with only ${\sim}7\%$. As can be expected, the single PS fit does perform notably worse in more cases $\left({\sim}22\%\right)$ than the single S\'ersic fit. Additionally, the number of cases where the single PS produces a significantly lower $\chi^2_\nu$ is only ${\sim}6\%$. This is because the single S\'ersic fit can attempt to account for a central point-source by increasing the S\'ersic index, whereas the PS model is limited in its ability to account for the host galaxy. A somewhat surprising result seen in both \textit{HST} and Subaru fits is that, in the vast majority of cases, the $\chi^2_\nu$ varies very little between the fits. A more in-depth discussion on the importance of $\chi^2_\nu$ and its relation to the fit quality is given in Section \ref{sec:discussion}.

When looking at the distribution of the ``good'' fits and those which were cut, we find the distributions of \textit{HST} and Subaru contain very similar features, despite the large difference in the point-source fraction distribution. Comparing the single S\'ersic \textit{HST} fits to the S\'ersic+PSF, we find that far more fits are cut from the region of point-source dominated sources that perform much worse in the single S\'ersic fit. For the single PS and S\'ersic+PS comparison, we again find that significantly more fits are cut from the region of point-source dominated fits, where the single PS fit performs equal to or better than the S\'ersic+PS fit. This result is due to our sample of sources and the selection of our cuts. Far more sources are likely to appear point-like for numerous reasons, such as the host galaxy having small enough angular size, the AGN completely dominating its host, or the extended wings of the galaxy falling below the limiting depth of the imaging. The minimum radius condition probes for all of these situations simultaneously, so it is expected that these sources are cut. Similarly, we expect to see this relation for the Subaru S\'ersic+PS cuts and their respective point-source fraction. What we find, however, is that the cuts see a similar relation with the $\chi^2_\nu$ rather than the point-source fraction. This suggests that there is a disconnect between these three properties for the Subaru fits, and so determining quality of fits becomes more complicated than for \textit{HST}. If we are unable to accurately determine the point-source fraction for Subaru sources, then determining the S\'ersic index becomes increasingly difficult.

\section{Investigating the Fitting Inconsistencies} \label{sec:discussion}

While Section \ref{sec:results} shows that we achieve reasonable agreement for the majority of parameters, the question of why the S\'ersic index fails in so many cases is important. If we are to completely understand AGN host galaxies, and even understand the AGN itself, an accurate quantitative indicator of galaxy morphology, such as the S\'ersic index, is needed. A first step into finding where the fits go wrong is by investigating the agreement between parameters as a function of redshift.

\begin{figure*}[t]
\centering
\includegraphics[width=\textwidth]{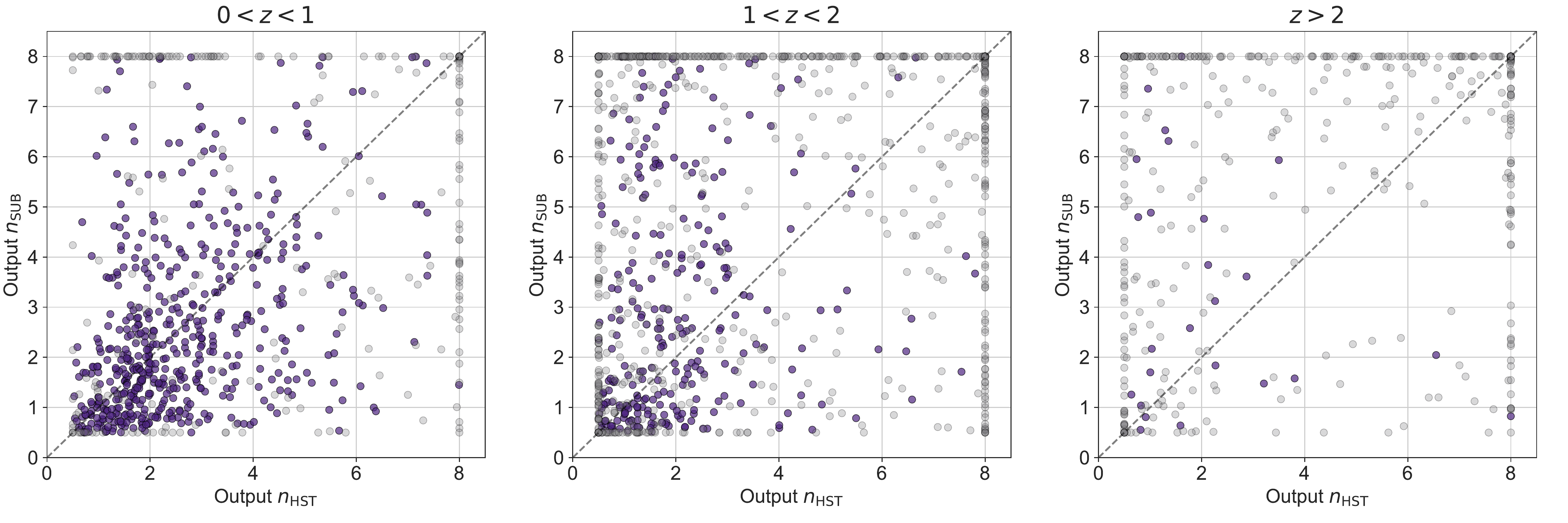}
\caption{Output S\'ersic index of Subaru vs. \textit{HST} distributed into redshift bins. The points marked in gray include all fits for both \textit{HST} and Subaru, regardless of whether the fit failed to pass the parameter/constraint cuts. The points marked in purple are those which are flagged as ``good'' for both \textit{HST} and Subaru. The 2951 sources are distributed into three redshift bins. The $0<z\le1$ bin contains 808 sources (of which 529 pass the parameter/constraint cuts), the $1<z\le2$ bin contains 1291 sources (263 after cuts), and the $z>2$ bin contains 852 (25 after cuts) As redshift increases, far more fits are cut. However, even at lower redshifts we do not see strong correlation between the \textit{HST} and Subaru S\'ersic indices.}
\label{fig:redshift_sersic}
\end{figure*}

We find that both the host and point-source magnitudes tend to agree regardless of redshift. There are a small number of faint sources in which the magnitudes disagree. These faint sources naturally tend to be at higher redshift, but for the vast majority of sources we saw no redshift dependence. The agreement of the effective radius begins to see some dependence on the redshift. This is expected, as the more distant a source is, the smaller its angular size. This angular size can reach values below the width of the PSF, and thus the effective radius of the source cannot be accurately determined as the source appears point-like. Figure \ref{fig:redshift_sersic} shows the output S\'ersic index for Subaru versus \textit{HST} separated into three redshift bins. We find that the agreement between the S\'ersic index does see some dependence on the redshift. As the redshift increases, we find that more sources tend to fail the parameter/constraint cuts, most commonly the minimum radius cutoff and the S\'ersic index constraint. In turn, we find many more ``good'' fits at lower redshifts. The lowest redshift bin $\left( 0<z<1 \right)$ sees 10\% and 15\% of sources flag the S\'ersic constraint for \textit{HST} and Subaru fits, respectively. The  $1<z<2$ bin has 28\% and 34\%, and the final $z>2$ bin sees 37\% and 41\%. Interestingly, while the fits at lower redshifts safely converge, there is still very little agreement between \textit{HST} and Subaru.

\begin{figure*}[t]
\centering
\includegraphics[width=\textwidth]{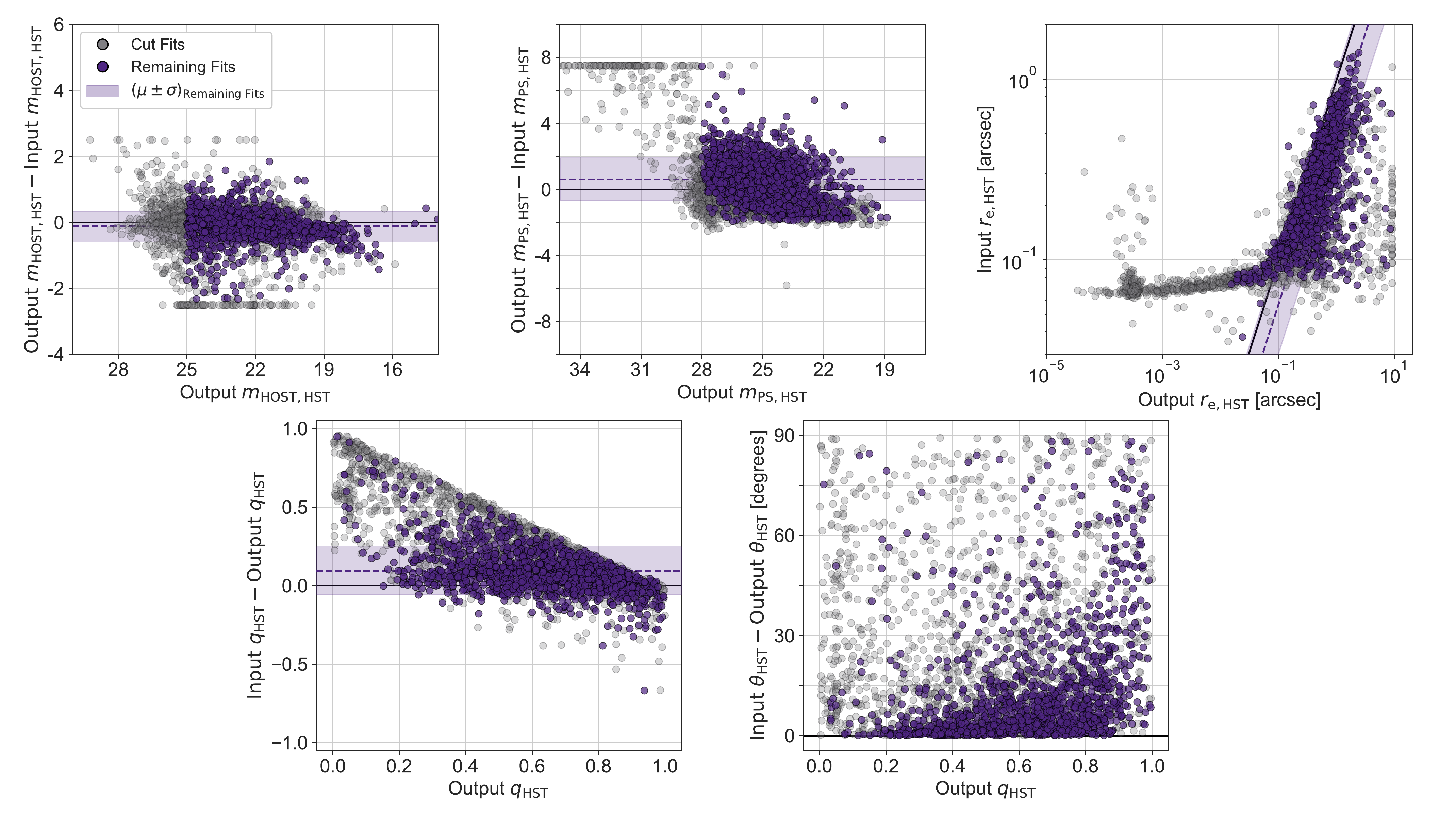}
\caption{Comparison of the \textit{HST} S\'ersic+PS input parameters determined from Source Extractor and the output from GALFIT. The points marked in gray include all fits, regardless of whether the fit failed to pass the parameter/constraint cuts. The points marked in purple are those which are flagged as ``good''. The solid, black line represents agreement between the input and output. The purple region represents the average difference between the two.}
\label{fig:in_vs_out_hst}
\end{figure*}

Another factor which could cause the disagreement between fits are the initial conditions and how they relate to the final output. Figure \ref{fig:in_vs_out_hst} compares the initial conditions determined through Source Extractor to the output of the \textit{HST} S\'ersic+PS fits. We can clearly see reasonable agreement between the host and point-source magnitudes until $m_\mathrm{out}\sim28$. The PS magnitude sees significantly more scatter than that of the host magnitudes. This is due to the input PS magnitude being defined in terms of the host magnitude rather than some measured property of the source. For both magnitudes we find that the regions of significant difference are caught by the parameter/constraint cuts. The axis ratio and position angle both see significant agreement between the input and output. Notably, a large number of cut fits have significantly higher input axis ratio than the output. These sources are the point-like sources caught by the minimum radius cutoff and S\'ersic index constraint. The difference comes from how Source Extractor and GALFIT treat these sources. In general, Source Extractor will list the axis ratio of a point-like source as $q\sim1$, whereas the axis ratio provided by GALFIT will wander throughout the fitting process and can end up at any point $0<q<1$ with no significant effect on the other fit parameters. For the effective radius, we see that there is reasonable agreement above a certain input radius. This radius is likely a minimum-allowed value in Source Extractor and corresponds to roughly 3 pixels. The vast majority of fits near this value are cut via the minimum radius cutoff and S\'ersic constraint. This collection of sources makes up the majority of the total cut fits. Of the ``good'' fits, we find that Source Extractor systematically underestimates the radius compared to GALFIT. This discrepancy in radius measurement between GALFIT and Source Extractor is well established and has little effect on the final parameters \citep{Haussler2007}. On average, we find that the input radii are $0.58 \pm 1.88$ times smaller (measured in log-space) than the output.

\begin{figure*}[t]
\centering
\includegraphics[width=\textwidth]{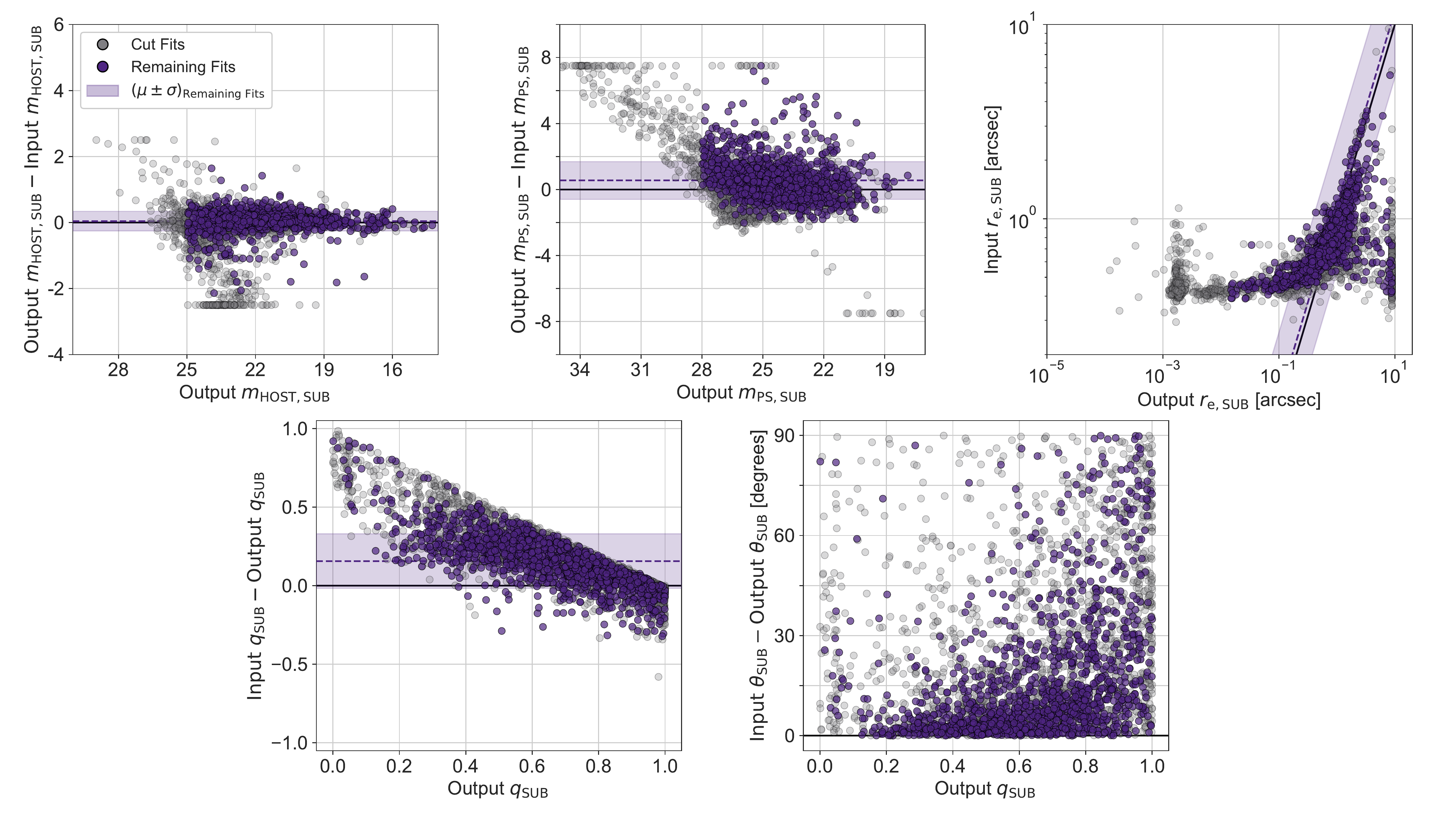}
\caption{Same as Figure \ref{fig:in_vs_out_hst} but for Subaru S\'ersic+PS fits.}
\label{fig:in_vs_out_sub}
\end{figure*}

Figure \ref{fig:in_vs_out_sub} shows the same results as Figure \ref{fig:in_vs_out_hst} but for the Subaru S\'ersic+PS fits. The results are similar, however the agreement in each case is worse. The host magnitude only sees agreement out to $m_\mathrm{out}\sim26.5$. A number of fits between $25<m_\mathrm{out}<22$ see significant disagreement between the input and output, however most are cut. With respect to the PS magnitudes, we similarly see agreement only until $m_\mathrm{out}\sim27$ with a higher number of fits reaching magnitudes dimmer than 28. Within the ``good'' fits, we see similar levels of agreement in the input and output magnitudes for both \textit{HST} and Subaru. There is a strong disagreement between the input and output axis ratio, however this did not cause any significant difference in any of the output parameters. Despite this, the Subaru input and output position angles see similar levels of agreement as \textit{HST}. The initial radii again bottom out to a minimum value corresponding to roughly 3 pixels (note that this minimum is higher than the minimum for \textit{HST} due to the difference in pixel scale). For Subaru, some of the fits near this value are not cut out, as they still provide reasonable agreement with \textit{HST}. Again, Source Extractor underestimates the radius of most ``good'' fits compared to the GALFIT output, however to a lower degree. Additionally, the difference measurement visualized in Figure \ref{fig:in_vs_out_sub} is offset due to the number of fits near the minimum input radius that were not cut.

The majority of the $n>5$ fits do not pass the parameter/constraint cuts -- most reach the maximum S\'ersic value of $n=8$ or fall below the $r_\mathrm{e}=0.015"$ cutoff and so may not provide the highest quality fits. There exist very few fits near $n=8$ with larger radii. The discrepancy between the S\'ersic indices, most notable at higher indices, is in large part due to the lower Subaru data angular resolution. Section \ref{sec:testing} shows that, while seeing the poorest agreement of all parameters, we achieved reasonably consistent agreement between the \textit{HST} S\'ersic index and the work of \citet{Gabor2009}. This result, alongside the expected nature of the S\'ersic index seen in the \textit{HST} portion of Figure \ref{fig:sersic_hist}, indicates that it is likely the Subaru fits being unable to accurately determine the S\'ersic index rather than a fault with the \textit{HST} fits.

\begin{figure*}[t]
\centering
\includegraphics[width=\textwidth]{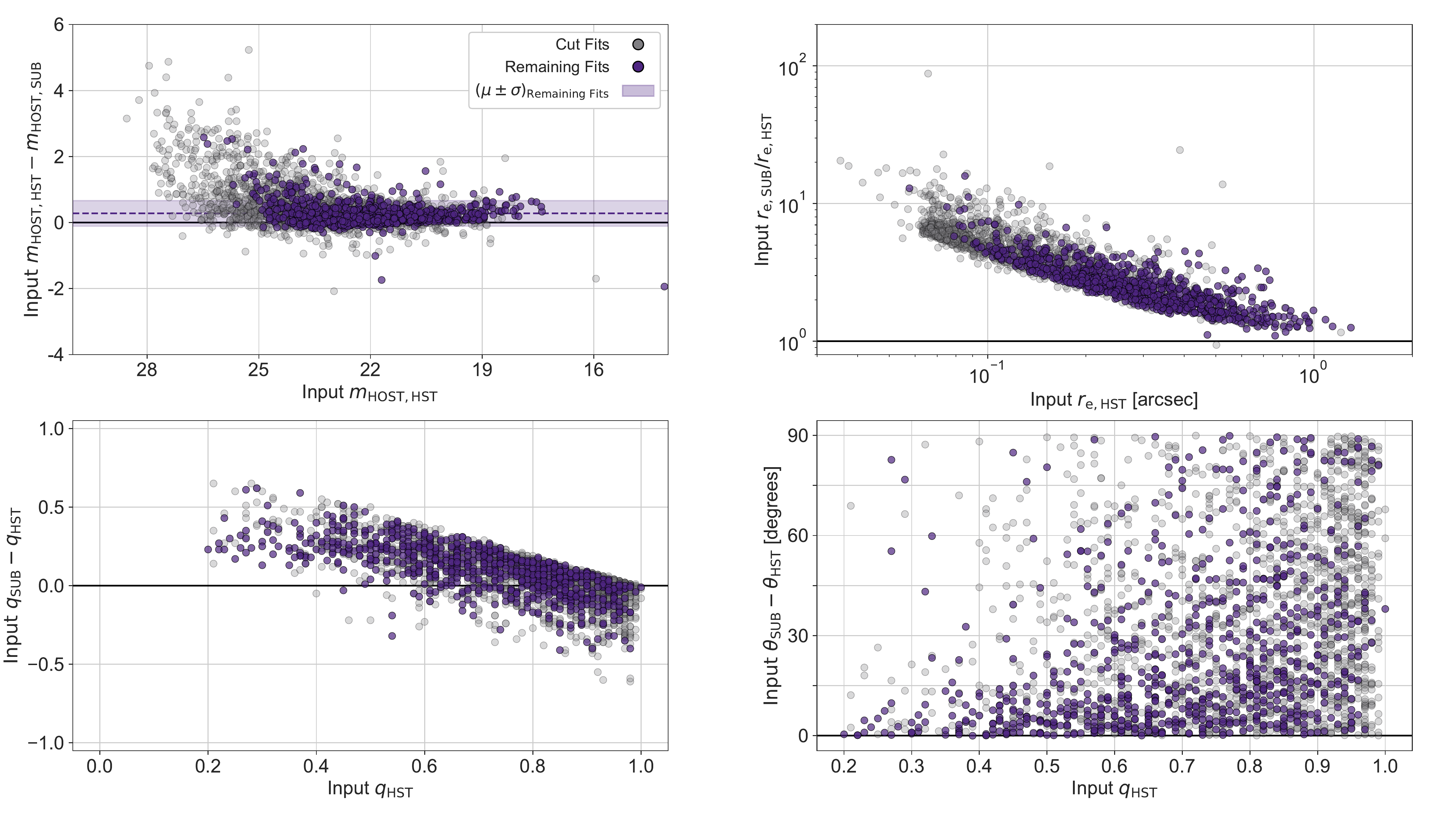}
\caption{Comparison of the \textit{HST} and Subaru S\'ersic+PS input parameters determined from Source Extractor. The points marked in gray include all fits for both \textit{HST} and Subaru, regardless of whether the fit failed to pass the parameter/constraint cuts. The points marked in purple are those which are flagged as ``good'' for both \textit{HST} and Subaru. The solid, black line represents agreement between the input and output. The purple region in the upper-left plot represents the average difference between the two. The other plots see no strong agreement.}
\label{fig:in_vs_in}
\end{figure*}

Figure \ref{fig:in_vs_in} compares the difference in the input parameters of \textit{HST} and Subaru in order to contextualize the effect of the choice of the initial values on the output of the fits. We see that the host magnitudes (and, by definition, the PS magnitudes) see significant agreement for sources brighter than $m_\mathrm{HST} \sim 25$, with the ``good'' Subaru fits' inputs averaging $0.27 \pm 0.39$ magnitudes brighter than \textit{HST}. This aligns with what we see with the parameter outputs in Figure \ref{fig:out_vs_out_sub}. The input radii, however, see very little agreement between \textit{HST} and Subaru. This is not necessarily unexpected given the results of Figures \ref{fig:in_vs_out_hst} and \ref{fig:in_vs_out_sub}, as there are distinct regions where the disagreement forms. The input radii reaches its minimum of ${\sim}3$ pixels for both \textit{HST} and Subaru, however these correspond to different angular sizes leading to significant disagreement for the lowest input radii. For radii above this point, we find that Source Extractor underestimates the radius of \textit{HST} sources relative to GALFIT more significantly than for Subaru, thus leading to the large difference at higher radii. The input axis ratio and position angles also see significant disagreement relative to the outputs seen in Figure \ref{fig:out_vs_out_sub}, although disagreement to a certain degree is expected, as discussed in Section \ref{subsec:parameters}. Despite the disagreement seen in a number of the input parameters, there exists no trend explaining the failure of the S\'ersic index.

The lack of correlation between the input parameters and the disagreement in the S\'ersic index implies that it may not be a fault of the fitting process, but perhaps of the $\chi^2_\nu$ minimization method used by GALFIT. As seen in Figure \ref{fig:chi2nu_results_sub}, many sources achieved very similar $\chi^2_\nu$ values in each of the three types of fits. In order to ensure that the solution output by GALFIT is the ``true'' solution, and not perhaps the algorithm being caught in a local minimum causing the disagreement between \textit{HST} and Subaru, we create a large number of models for a single source covering a large portion of the morphological parameter space for both \textit{HST} and Subaru. This allows us to investigate how the $\chi^2_\nu$ changes with certain parameters in order to understand where the fits begin to fail. Notably, this method is independent of the fitting process and only depends on GALFIT's $\chi^2_\nu$ calculation.

For each source we investigated in this manner, we created a 4D grid of the host magnitude, PS magnitude, radius, and S\'ersic index for both \textit{HST} and Subaru. Using each point in this grid, we use GALFIT to create a model and calculate the $\chi^2_\nu$. Alongside this grid, we used the best fit values for the remaining parameters, namely the host and PS positions, position angle, and axis ratio. The radius grid was defined as 20 points varying logarithmically from $10^{-4}$--$10^{1}$ arcsec. This range was selected as these values were the minimum and maximum values seen throughout the fitting process, despite the minimum range clearly being nonphysical at our angular resolution. This was performed in order to investigate the cause of the fits congregating at very low radii and thus we include this large range. We then selected 20 points along the entire range of allowed S\'ersic indices, namely from 0.5 to 8. The host and PS magnitude axes only include 5 points. This, alongside the range of magnitudes, was selected somewhat subjectively. We found little significance in the changes between points less than 1 magnitude apart, so the grid spacing was defined as 1 magnitude. We also found that, for most galaxies, a selection of 5 grid points was sufficient in order to visualize the entire parameter space. In most cases, we center the magnitude grids on the best fit values rounded to the nearest half-magnitude. In certain cases, such as those fits which find very dim (${\sim}30$ or fainter) PS magnitudes, this grid is defined with the best fit as the minimum point in the grid as opposed to the center. The grid spacing for the PS magnitudes is set instead to 2 magnitudes in many of these cases in order to investigate beyond the faint PS domain.

\begin{figure*}[p!]
\centering
\includegraphics[width=\textwidth]{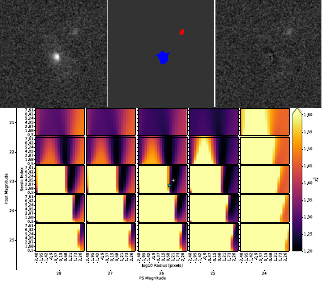}
\caption{The reported $\chi^2_\nu$ as measured for all $10^4$ models for source cid\_1010 imaged by \textit{HST}. The \textit{HST} cutout (upper left), pixel mask (upper middle) and best fit residual (upper right) are also given. In the pixel mask, the blue source is the primary source to be fit and the red sources are masked out in GALFIT and the $\chi^2_\nu$ calculation. The white star represents the best fit and the cyan star represents the initial conditions (with magnitudes rounded to the nearest point on the grid). Each plot in the grid represents the $\chi^2_\nu$ for varying radius and S\'ersic index, while each plot along the major axes represents how the parameter space varies with changing magnitude. The hue of the heatmaps indicates the $\chi^2_\nu$ of the corresponding model. The minimum value in the colourmap is defined as the best fit $\chi^2_\nu$. Note that the $\chi^2_\nu$ extends beyond the maximum hue as the $\chi^2_\nu$ quickly increases for very poor fits. See Figure \ref{fig:cid_1010_sub} for the corresponding Subaru models.}
\label{fig:cid_1010_hst}
\end{figure*}

\begin{figure*}[p!]
\centering
\includegraphics[width=\textwidth]{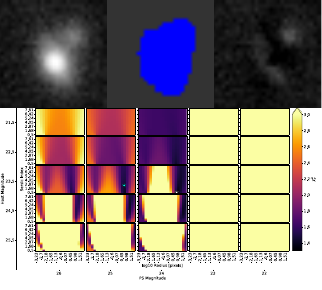}
\caption{The $\chi^2_\nu$ as measured for all $10^4$ models for source cid\_1010 imaged by Subaru. See Figure \ref{fig:cid_1010_hst} for the \textit{HST} equivalent and a description of the figure. Note the different grid points and color range between the two figures. The blue source in the pixel mask is the main source to be fit. Note that Source Extractor could not distinguish the source nearest the main source as being separate, and thus GALFIT treats them as a single source.}
\label{fig:cid_1010_sub}
\end{figure*}

Using this 4D grid, we achieve $10^4$ different combinations of parameters. Using GALFIT, we created a model and calculated the $\chi^2_\nu$ for each combination. Note that we do not use GALFIT to fit the source; we simply create a model using the parameters. We are then left with a 5D system representing the parameter space over which we can investigate how the quality of fit varies. Figures \ref{fig:cid_1010_hst} and \ref{fig:cid_1010_sub} give examples of this visual representation for a typical source. The displayed source was selected such that both the \textit{HST} and Subaru were flagged as ``good'' fits. In particular, this source is shown as it represents many of the situations that can cause a fit to go wrong while still representing the characteristic parameter space for both \textit{HST} and Subaru and still being flagged as ``good''. For example, this source features a faint extended disk which, while more obvious in the Subaru imaging, goes relatively undetected in the \textit{HST} fit. In addition, Source Extractor was unable to distinguish between the primary source and its neighbor in the Subaru image, thus an accurate measurement of the host's morphology is impossible. Both of these factors increase the effective radius reported by Source Extractor, thus increasing the size of the pixel mask -- thus more distant neighbors are within the range to be fit rather than be masked out. Despite this, this source is still representative of the typical parameter space. Looking at the \textit{HST} residual image given in Figure \ref{fig:cid_1010_hst}, we clearly see that the best fit is a relatively good fit as there are no clear regions within the source where the fit failed. The Subaru fit seen in Figure \ref{fig:cid_1010_sub}, however, is of lower quality due to the misidentification of the primary source and its neighbor. This resulted in the best fit parameters not agreeing. Despite this, the distributions of $\chi^2_\nu$ between the two fits actually take very similar form. It is worth noting that, while the features exist in similar qualitative regions of the parameter space (e.g., faint point-source magnitude, bright host magnitude), they do not agree quantitatively. For example, the third column in Figure \ref{fig:cid_1010_hst} is visually very similar to the third column in Figure \ref{fig:cid_1010_sub}, but these columns correspond to a point-source magnitude 2 magnitudes apart (26 and 24 for \textit{HST} and Subaru, respectively).

Throughout all the visualizations, including those displayed in Figures \ref{fig:cid_1010_hst} and \ref{fig:cid_1010_sub}, we see similar features that are able to explain many of the trends which we saw throughout Section \ref{sec:results}. Varying the host magnitude has a relatively strong influence on the $\chi^2_\nu$ of the fit. This, in part, explains why the host magnitude agrees in the majority of cases. Varying the PS magnitude, however, has a lower influence on the $\chi^2_\nu$, though still having an impact. In general, the effect of the PS magnitude is low until a certain point at which the point-source is too bright and the $\chi^2_\nu$ quickly increases. This likely explains that, while there is reasonable agreement between \textit{HST} and Subaru, we see the wider scatter in the upper-middle panel of Figure \ref{fig:out_vs_out_sub}. In many of the radius versus S\'ersic index subplots in Figures \ref{fig:cid_1010_hst} and \ref{fig:cid_1010_sub}, we tend to see two groups of radii which provide a low $\chi^2_\nu$. There is typically a region at higher radii where the best fit lies, but there is also a grouping at very low radii seen in both \textit{HST} and Subaru. These radii correspond roughly to the same value as the smallest radii fits seen in Figure \ref{fig:out_vs_out_sub}. This implies that those groups of fits are caught in this low radius region and may have a more suitable fit at higher radii, despite the low radius having a lower $\chi^2_\nu$. Figure \ref{fig:cid_1010_hst} is somewhat atypical in the case of the low radius grouping as, in the majority of cases, the \textit{HST} visualizations have larger groupings more similar to those seen in Figure \ref{fig:cid_1010_sub}. It is likely that the radius of this \textit{HST} fit is more well-constrained than in many other cases.

\begin{figure*}[t]
\centering
\includegraphics[width=\textwidth]{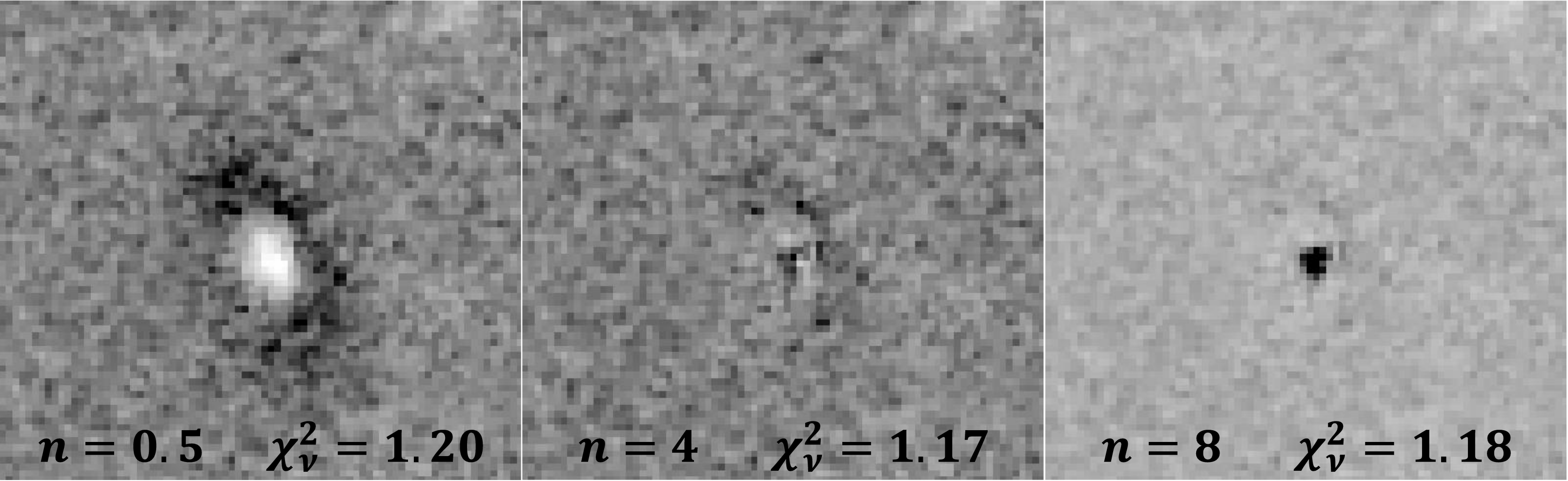}
\caption{Comparison between the residuals of three models for the same \textit{HST} source, namely cid\_1010. All three models use the best fit values for each parameter other than the S\'ersic index. The reported $\chi^2_\nu$ of each model is also provided.}
\label{fig:cid_1010_hst_sersic_comp}
\end{figure*}

Figures \ref{fig:cid_1010_hst} and \ref{fig:cid_1010_sub}, as well as the other visualizations not displayed, also provide an interesting perspective into the S\'ersic index of these fits. In all of these plots we see that varying the S\'ersic index has a much smaller influence on the $\chi^2_\nu$, as the features all contain long, vertical lines of similarly valued $\chi^2_\nu$. Figure \ref{fig:cid_1010_hst_sersic_comp} shows an example of this for the same \textit{HST} source as Figure \ref{fig:cid_1010_hst}. The three fits shown each have drastically different residuals, with the $n=0.5$ model clearly failing both at the central point-source as well as for the extended galaxy and the $n=8$ heavily over-fitting the bulge. The best model is the $n=4$ model; however, the $\chi^2_\nu$ of each is very similar. While the $n=4$ model does have the lowest $\chi^2_\nu$, the fact that the $\chi^2_\nu$ changes so little for such clear failings in the model is concerning. The limited effect of changes in S\'ersic index on $\chi^2_\nu$ we see here likely results in the disagreement seen in Figures \ref{fig:Gabor_Comparison}, \ref{fig:out_vs_out_sub}, and \ref{fig:out_vs_out_conv}.

These results indicate that $\chi^2_\nu$ alone is not a good indicator of quality of fit. A visual inspection of each fit is required to truly measure the quality of fit. In addition, a simultaneous investigation of the morphological parameters is necessary to ensure they remain physically acceptable. For example, a fit may have a low $\chi^2_\nu$ with a good residual image, however the same fit may have a radius of 0.01 pixels with an axis ratio of $10^{-4}$. This result is clearly not physical despite providing a good residual and $\chi^2_\nu$. In many cases, these unrealistic parameters are not visible in a residual image as they are typically faint relative to the rest of the source. To prevent this, stricter parameter constraints should be applied to these parameters.

Within Figures \ref{fig:cid_1010_hst} and \ref{fig:cid_1010_sub}, we are also able to reveal underlying relations between the host magnitude, PS magnitude, S\'ersic index, radius, and $\chi^2_\nu$. As the host magnitude becomes fainter and the PS magnitude becomes brighter, the radius must increase in order to remain at a low $\chi^2_\nu$. The S\'ersic index tends to increase along with the radius, resulting in an upward-diagonal feature of low $\chi^2_\nu$ along the S\'ersic index-radius space. A similar feature in the opposite direction is also noticeable for the lower-radius grouping, where we see the radius decrease with decreasing host magnitude and PS magnitude. In addition, the lower-radius grouping tends to see its S\'ersic index decrease as the radius decreases.

\section{Conclusion}

We applied GALFIT to calculate the morphological parameters of a set of 4016 X-ray AGN in the COSMOS field spanning redshifts $0.03 \lesssim z \lesssim 6.5$ using \textit{i}-band imaging from both \textit{HST} and Subaru. We performed three fits to each source: a single S\'ersic profile fit, a single PS fit, and a two-component S\'ersic+PS fit. After testing our method against the work of \citet{Gabor2009} to ensure that our method could consistently extract the morphological parameters from an \textit{HST} source, we compared the results of the \textit{HST} fits to those of the Subaru fits.

We found that there was strong agreement in a number of morphological components, notably the host galaxy magnitude, the PS magnitude, and the host galaxy effective radius. Importantly, the S\'ersic index saw virtually no agreement between the different sets of fits. This disagreement seems to be completely independent of the other morphological parameters of the fit. There is also a relation between when the S\'ersic indices disagree and the redshift of the source. We find that many more sources fail to converge within the parameter constraints at higher redshifts, but still see virtually no agreement for the S\'ersic index even at low redshifts. By testing our method against that of \citet{Gabor2009}, we were able to recover morphological parameters consistent by those reported by \citet{Gabor2009}, including for the S\'ersic index. In addition to this, Figure \ref{fig:sersic_hist} follows the expected relation between the S\'ersic index and the addition of a PS component to a galaxy fit for \textit{HST} sources; namely, if one does not account for the central point-source, the S\'ersic index will systematically increase. For Subaru sources, however, this relationship was less significant. The results of Figure \ref{fig:chi2nu_results_sub} also acts as a validation of the method; for \textit{HST}, we see an intuitive relation between the $\chi^2_\nu$, the point-source fraction, and the parameter/constraint cuts, whereas there exists a disconnect of these parameters for Subaru sources. These tests of the method increases our confidence that the \textit{HST} fits are able to retrieve the S\'ersic index relatively consistently, thus implying that the lower resolution data are the likely cause of the disagreement.

In order to determine whether the resolution difference was the primary cause of the disagreement, we created a new set of data by convolving the \textit{HST} cutouts with the Subaru PSF. We applied our fitting process to this new set of data and found similar results to that of the Subaru fits, with strong agreement between most parameters, but none between the S\'ersic index. After comparing the convolved \textit{HST} results to that of Subaru, we found that the S\'ersic index still did not agree even between the two low resolution sets of data. This test seems to confirm that the S\'ersic index tends to fail for these low resolution, broad PSF fits. Since we are attempting to fit galaxies with such small angular size, the galaxy itself is comparable in size to the PSF of the telescope, thus determining accurate morphologies when AGN are present is difficult.

As GALFIT performs its fits by minimizing $\chi^2_\nu$ for the model, we attempted to investigate how the fits fail by viewing the relationship between $\chi^2_\nu$ and the morphological parameter space for a large number of sources, a representative example of which is given in Figures \ref{fig:cid_1010_hst} and \ref{fig:cid_1010_sub}. We found that the parameters that tended to agree most strongly (i.e., host magnitude and radius) had the most significant effect on the $\chi^2_\nu$, and conversely the S\'ersic index had little effect on $\chi^2_\nu$ over its entire range. This is likely an additional cause of the disagreement between the \textit{HST} and Subaru or convolved \textit{HST} fits. This result shows that $\chi^2_\nu$ alone is not necessarily a gauge of quality of fit. The residual must also be viewed in order to ensure that the model is fitting as intended. This viewing also ensures that features which cannot be accounted for have no significant influence on the fit parameters, for example an overlapping source or spiral arms.

Small areas of low $\chi^2_\nu$ also appeared consistently within the parameter space at non-physical values. These non-physical values include extremely low radii, high S\'ersic indices, and extreme axis ratios. These values may not necessarily show in the image residual in certain cases, for example in a point-like galaxy. Thus, in order to keep parameters physical, constraints to prevent unrealistically low radii or axis ratios must be applied. In addition, simultaneous comparison of the residual image alongside viewing the morphological parameters is required to fully judge the quality of a fit.

It is possible that a different fitting software using a different algorithm, or perhaps an alternative technique entirely, may be able to more accurately determine the S\'ersic index. Neural networks are being developed to perform studies similar to this without fitting a surface brightness profile \citep{Ghosh2020}. Further work in comparing the morphologies determined through the use of neural networks to those determined through surface brightness profiles is required to determine if either, or perhaps both used in conjunction, may provide a more well-constrained morphological classification. Overall, we recommend that GALFIT-derived S\'ersic indices for high-redshift, low angular size, active galaxies imaged at lower angular resolution are interpreted with caution until further studies are able to constrain the morphology. Future work includes simulating a similar sample of galaxies at a wider range of redshift, signal-to-noise, angular size, etc. for imaging spanning a wider range of telescope resolutions and depths to determine a more quantitative understanding of when the S\'ersic index is reliable. Performing fits similar to those in this study would allow for a more rigorous statistical analysis of how well the AGN and its host are disentangled and whether we can determine the morphology. Additionally, we would have the ability to better constrain the uncertainties of the morphological parameters. These analyses in combination with the $\chi^2_\nu$ mapping of Section \ref{sec:discussion} would allow for a more rigorous test than previous studies.

\begin{acknowledgments}
We would like to thank the anonymous referee for their thorough review of the manuscript. The suggested changes greatly improved the quality of this work.

C.D., P.B., and S.C.G. acknowledge the support of the Natural Sciences and Engineering Research Council of Canada (NSERC). Nous remercions le Conseil de recherches en sciences naturelles et en génie du Canada (CRSNG) de son soutien.

C.D. thanks the Ontario Graduate Scholarship (OGS) for their support throughout this project.

C.M.U. acknowledges support from the National Science Foundation under Grant No. AST-1715512.

Thank you to Dave Sanders, Jane Turner, Nico Cappelluti, and the rest of the Accretion History of AGN collaboration (\url{https://project.ifa.hawaii.edu/aha/}) for their feedback and support over the course of this project, including the supplying of the \textit{HST} cutouts and PSF.

The Hyper Suprime-Cam (HSC) collaboration includes the astronomical communities of Japan and Taiwan, and Princeton University. The HSC instrumentation and software were developed by the National Astronomical Observatory of Japan (NAOJ), the Kavli Institute for the Physics and Mathematics of the Universe (Kavli IPMU), the University of Tokyo, the High Energy Accelerator Research Organization (KEK), the Academia Sinica Institute for Astronomy and Astrophysics in Taiwan (ASIAA), and Princeton University. Funding was contributed by the FIRST program from the Japanese Cabinet Office, the Ministry of Education, Culture, Sports, Science and Technology (MEXT), the Japan Society for the Promotion of Science (JSPS), Japan Science and Technology Agency (JST), the Toray Science Foundation, NAOJ, Kavli IPMU, KEK, ASIAA, and Princeton University.

This paper makes use of software developed for Vera C. Rubin Observatory. We thank the Rubin Observatory for making their code available as free software at \url{http://pipelines.lsst.io/}.

This paper is based on data collected at the Subaru Telescope and retrieved from the HSC data archive system, which is operated by the Subaru Telescope and Astronomy Data Center (ADC) at NAOJ. Data analysis was in part carried out with the cooperation of Center for Computational Astrophysics (CfCA), NAOJ. We are honored and grateful for the opportunity of observing the Universe from Maunakea, which has the cultural, historical and natural significance in Hawaii.
\end{acknowledgments}

\vspace{5mm}

\facilities{\textit{HST} (ACS),
            Subaru (HSC)
}
\software{TinyTim \citep{Krist2011},
          GALFIT \citep{Peng2002, Peng2010},
          Source Extractor \citep{Bertin1996},
          Montage \citep{Jacob2010},
          Astropy \citep{AstropyCollaboration2013,AstropyCollaboration2018}
}

\bibliography{bibliography}{}
\bibliographystyle{aasjournal}

\end{document}